# MORTAL COMPUTATION: A FOUNDATION FOR BIOMIMETIC INTELLIGENCE

**Alexander Ororbia**
Rochester Institute of Technology
Rochester, NY 14623
ago@cs.rit.edu

**Karl Friston**
VERSES AI Research Lab
Los Angeles, CA 90016, USA
karl.friston@verses.ai

## ABSTRACT

This review motivates and synthesizes research efforts in neuroscience-inspired artificial intelligence and biomimetic computing in terms of *mortal computation*. Specifically, we characterize the notion of mortality by recasting ideas in biophysics, cybernetics, and cognitive science in terms of a theoretical foundation for sentient behavior. We frame the mortal computation thesis through the Markov blanket formalism and the circular causality entailed by inference, learning, and selection. The ensuing framework — underwritten by the free energy principle — could prove useful for guiding the construction of unconventional connectionist computational systems, neuromorphic intelligence, and chimeric agents, including sentient organoids, which stand to revolutionize the long-term future of embodied, enactive artificial intelligence and cognition research.

## 1   Introduction

*Life can only be understood backwards; but it must be lived forwards.*
-Søren Kierkegaard [195]

A key aspect of biomimetics is that of biomimicry or, rather, emulating natural or biological systems, when crafting mathematical, computational processes or technological artifacts. This is done generally with the aim of solving a complex problem, potentially facilitating bionics — and the engineering of a nature-oriented system — for real-world applications. It follows that biomimetic and bionic intelligence research seeks to construct the cognitive functionality and processes that underwrite intelligence by imitating or emulating natural minds and brains, the products of countless centuries of evolution.

Life itself serves as an unceasingly fecund source of inspiration, capturing the attention of countless theoreticians and engineers; ranging from biology to physics — to philosophy to computer science. Living systems, from single-cell microorganisms to complex creatures such as plants and animals, offer a remarkable example of how homeorhesis (and related notions of allostasis and autopoiesis) — the self-regulating processes employed by biological systems to assemble and maintain themselves [71, 72, 53] — can support the emergence of adaptive and complex behavior. From the computational point-of-view, homeorhesis shapes and directs the 'calculations' conducted by the organism as it interacts with its environment; specifically, its ability to infer, learn, and evolve. From the standpoint of cybernetics, emulating the computational properties of living systems goes far back, notably within the framework of self-reproducing cellular automata [345, 148, 335, 359].

Our central argument is that if the aim is to craft mechanistic, computational systems capable of the vast array of animal or human-like behaviors, the thrust of machine intelligence will need to change over the coming decades, similar in spirit to the call for a shift from symbolic to non-symbolic processing [107, 106, 26], which motivated classical connectionism and statistical learning. Machine intelligence research may need to focus on the processes that realize its nonlinear self-organization and efficient adaptation; such a move is towards the development of survival-oriented processing that embodies computational notions of life/mortality, a sort of *naturalistic machine intelligence*. Specifically, we argue that a crucial pathway — on the way to artificial general intelligence (AGI) – lies in *mortal computing*, a framework that we seek to develop in this article, hoping to engage researchers pursuing the sciences of the artificial [311].

At this point, one might ask: *what is mortal computation*? 'Mortal computation', originating in [170] and further developed and advocated for in [262, 254], asserts that the mathematical calculations/processes that underlie information processing in a biological or artificial system are inseparable from the physical substrate that implements and executes them, i.e., the 'software' cannot be divorced from the 'hardware'. This stands in strong contrast with the notion of computation in Computer Science: here, software is decoupled from hardware and software is 'immortal', which means that it can be copied to different hardware and still be executable. Any machine learning algorithm, which can be viewed as program that adjusts itself in accordance with data, also relies upon this separation (and is designed with this in mind). Furthermore, immortal computation means that a program, even an adaptive one, e.g., an 'intelligent system', develops characteristics and acquires knowledge irrespective of the medium upon which it is instantiated. In contrast, mortal computation means that once the hardware medium upon which a program is implemented within fails, or 'dies', the knowledge, behavior, and specific functionality, including its 'quirks', will also cease to exist. This is much akin to what would happen to the knowledge/behavior acquired/developed by a biological organism once it is no longer able to maintain itself.

We argue that, at least for AGI, it may be necessary to consider mortal computation, if not entirely abandoning immortal computation. Transitioning to a kind of computation that is tightly bound to its substrate's fate might at first appear unnecessary or even foolhardy; given that that such programs could be brittle and lack the universality often sought in intelligent systems research. However, particularly when considering some of the pressing global challenges we face today, e.g., climate change, we will foreground the advantages that attend mortal computation — advantages which are difficult to realize in the context of immortal computation. Furthermore, efforts that embrace such a transition may find useful developments surprisingly easy. In general, the phrase mortal computation affords many perspectives on one central truth – *intelligent behavior, whether manifest in biological or artificial form, is inextricably intertwined with persistence in the face of an environment that could bring about its end*. This important notion unifies many ideas that have been articulated over the decades, across different disciplines, including philosophy, cybernetics, biophysics, computer science, cognitive science, and physiology, and has implications for a large swath of application domains in computational neuroscience, neuromorphic computing, and neurorobotics [79].

**Structure of the paper.** In this work, we describe a formal framework for mortal computation that may serve as a guide for addressing biomimetic systems, models for neural information processing, and AGI. First, in Section 2, we motivate the mortal computation thesis as necessary for biomimetics, drawing from thermodynamics, the notion of embodiment and enactivism, and the philosophy of life and death. Next, in Section 3, we will review historical concepts to characterize mortal computation in terms of biophysics, cybernetics, cognitive science and naturalist philosophy. In Section 4, we frame mortal computation from the perspective of *mortal inference, learning, and selection* based on the free energy principle, the Markov blanket formalism, and foundations of morphology. Section 5 offers a definition of mortal computation, elucidating several of its tenets that could feature in future developments in biomimetic and bionic intelligence. Finally, we will consider longer-term implications of the ensuing framework for scientific research in this direction.

## 2      Why Mortal Computation?

At first glance, there might appear to be a disadvantage in conflating some software's fate with that of the hardware that implements it: we are precluding the ability to duplicate/transfer our computer programs across platforms and the ability to design them with little worry as to how they will be executed. However, abandoning the immortal nature of computation brings with it something invaluable; namely, a substantial saving in energy usage and a reduction in the cost of creating the hardware needed to execute the requisite calculations. Furthermore, this shift sets the stage for a key way of thinking about artificial minds: intelligence and adaptive behavior are driven by an impetus to continue to persist in the face of an evolving environment or ecological niche (econiche). Notably, this framing of artificial intelligence (AI) reflects conceptual shifts seen over past decades in cognitive science and naturalist philosophy of mind.

**Thermodynamic Motivations.** To understand why energy savings comes naturally with a mortal form of computing, it is important to consider the thermodynamics of contemporary computing. In the context of open systems, there is an intimate relationship between information processing and thermodynamic efficiency as governed by Landauer's principle [208, 45] and the Jarzynski equality [186, 114, 14]. Specifically, Landauer's principle states that the minimum amount of energy $E$ required to erase a bit of information is proportional to the operating temperature $T$ of the (computing) system: $-E > k_B T \ln(2)$ where $k_B$ is Boltzmann's constant. Thus, an irreversible change in a computer's stored information requires the dissipation of a minimum level of heat to its surrounding environment. The Jarzynski equality, in effect, states that the difference of the free energies of states $X$ and $Y$ is constrained to be equal to the average over all the fluctuations in work done on all paths in taking that system from (equilibrium) state $X$ to (typically non-equilibrium) state $Y$, i.e.,

$$\exp\left(\frac{F_X - F_Y}{k_B T}\right) = \left\langle \exp\left(\frac{-W_{XY}}{k_B T}\right)\right\rangle$$

where $F_X$ is the free energy of state $X$, and $W_{XY}$ is the irreversible (stochastic) work done on the path taken from $X$ to $Y$, and $\langle \circ \rangle$ denotes averaging. Notably, the Jarzynski equality can be used to compute the Landauer limit independently of the speed of the irreversible logical procedure (e.g., bit erasure) under consideration [20]. Put simply, when Landauer's principle and the Jarzynski equality are taken together, there is a lower bound on the amount of thermodynamic work needed to change the information content of a computing system.

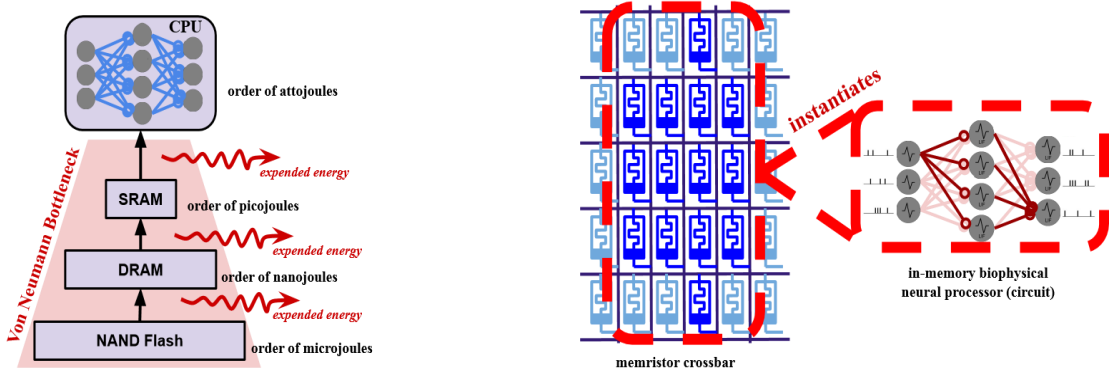

**Figure 1:** One of the key differences between immortal computation (Left), implementing a model such as the modern-day deep neural network (DNN), and mortal computation (Right), implementing a biophysical morphological model such as a spiking neural network, will be their thermodynamic costs. In a typical Von Neumann architecture that runs an immortal program, energy walls exist between different types of memory in the computer architecture, requiring an expenditure of energy to transfer information, e.g., synaptic weight values or intermediate layer activities, from non-volatile memory (NAND flash) to volatile memory of varying types (DRAM, SRAM) finally to the CPU housing a DNN. In a neuromorphic chip, such as one implemented as a memristor crossbar (Right), that would manifest a mortal program, processing happens directly on top of/within memory, circumventing the thermodynamic work required in a Von Neumann setup.

When viewed through the lens of the free energy principle [134], changing a computing system's information content can be read as inference (or belief-updating) in response to the external perturbations that a (e.g., neuronal) system is open to [137]. Furthermore, it can be shown that inference — cast as a variational free energy minimizing process — shares the same minima with thermodynamic free energy [305]. It then follows that the statistical and thermodynamic efficiency of any system — to which the free energy principle applies — are two sides of the same coin. This relationship, however, depends upon a description of systemic dynamics in terms of inference. In other words, the dynamics are interpreted as a gradient flow on variational free energy or, equivalently, (the logarithm of) Bayesian model evidence. This stands in contrast with the simulation of inference in modern-day computers, e.g., von Neumann architectures. In this simulation setting, there are additional thermodynamic costs that are induced by reading and writing to computer memory, i.e., the power consumption of moving data between off-chip memory units and a computing processor is nearly 100 times greater than the floating-point operations on those data. These costs are sometimes associated with the 'von Neumann bottleneck' or the 'memory wall' [368]. This means that, in order to realize the potential thermodynamic efficiency of Bayesian computations, it is necessary to implement belief updating in memory, also known as in-memory processing or processing-in-memory [154, 368].

From the perspective of machine intelligence (see Figure 1), these considerations license an important move to biomimetic computing based upon local [free] energy functionals and grounded in implicit in-memory computation, where memories correspond to the (c.f., synaptic) connection weights and the parameters of implicit generative models. From the perspective of mortal computation, these considerations speak to the importance of instantiating all aspects of computation in the substrate that will perform those computations. Broadly speaking, the calculations that characterize modern-day AI systems will, in effect, need to be brought as close as possible to the hardware that supports them; much as is the case in biological systems. Immortal computation, in contrast, will inherently require work to transfer the values of, for example, weight tensors to and from memory. It will further have to deal with random access memory, as the price paid will always relate to the read/write cost. The underlying physics of information necessarily dissolves any hard line between thermodynamic and statistical efficiency. In other words, to be Bayes optimal is to be thermodynamically efficient and *vice versa* [305, 306]. One can leverage this fact to furnish complementary perspectives on many aspects of neuronal architecture and physiology.

A general move to the thermodynamically efficient computing described above addresses concerns that have emerged in AI research [17, 62, 322, 303, 272]: how may we design systems that do not escalate computational and carbon costs? In essence, there is a need to move away from *Red AI*, which is dominated by immortal computation, towards *Green AI* [303], which mortal computation would naturally facilitate, speaking to global sustainable development goals [214, 240]. Neural transformers, which power modern of "generative AI", emit approximately 4.96 times the lifetime emissions of the average car [322][1], and, based on the number of graphics processing units (GPUs) shipped in 2022, it was estimated that the chat generative pre-trained transformer [286] (ChatGPT) produces 502 metric tons of carbon during its training (GPT-3 required 1287 megawatt hours) [232].

**Embodiment and Enactivism.** From the field of cognitive science, a core duality has emerged over the decades that echoes the divide between immortal and mortal computation. Specifically, it is between isolated versus situated or embodied cognition [128]. Isolated cognition takes the position that an entity without a 'body' could exhibit cognitive skills. It is possible to have a "brain in a vat" [284], i.e., a mind and its cognitive functionality is independent of its particular physical embedding or substrate; reminiscent of Descartes' mind-body dualism [29]. The theory of embodied cognition [355], in contrast, posits that the mind is grounded in the details of sensorimotor coupling: an organism's body intrinsically constrains and shapes the nature of its mental activity. This contrast between isolated and embodied cognition can be likened to arguing whether the software that underpins intelligence can exist without the hardware structure that manifests an agent in the physical world.

There is a great deal of evidence in support of embodiment, starting from the observation that humans reenact experiences based on the brain's modality-specific systems for perception and action. Representations of knowledge are influenced by previous interactions with the world and entail reactivation of sensorimotor functions in the absence of direct engagement with sensory input [9, 110, 5, 128]. Stronger variants of embodiment posit that the body serves a feedback-driven role in mental functioning: a situated agent can sense the world as well as be directly influenced by it, thus its behaviors do not require any representations or computational reference [37, 238, 353, 355]. The notion of mortal computation embraces the framing of embodiment, i.e., it views the physical morphology as integral to the nature of its computational processing. The central functions inherent to a mortal computer are hardware or 'body'-based and not sharply distinct from low-level sensorimotor functions. This arguably means that a model of mortal computing can be described as a particular instance of 'embodied intelligence' (or embodied AI [108]). However, as we will see in our treatment of mortal computation, such a system is more than embodied processing; it is also an incarnation of enactive functionality. In complement to embodiment, enactive cognition emphasizes the dynamic relationship between a living entity's environment and the entity itself [331]; an entity's cognition is the result of exercising its sensorimotor systems to exchange with its world. This means that a mortal computer is an active participant in the generation of the information that it processes and thus will be shaped by the consequences of how it acts and has acted [104, 342].

**The Philosophy of How Death Shapes Life.** Philosophically, the "finitude" of life can be viewed as one of the factors that endows life with purpose, creating the 'impetus' (e.g., selective pressure) for niche construction. This impetus drives us humans to communicate (e.g., teach) to pass on wisdom and heuristics from generation to generation; i.e., cultural niche construction, as seen through the lens of evolutionary psychology. This allows us to create works of art to express an inner message or embodied experience. Furthermore, life's finiteness entails reproduction, where genetic predispositions are passed on that ensure the survival of the species. This perspective motivates the self-replicating biophysical and cybernetical aspects of a mortal computer considered in Section 3. From the perspectives of philosophy of life/death [125] and existentialism [126], death — or the termination of an entity's existence — is a horizon that implicitly shapes behavior and consciousness. In other words, death is what allows future events to exert an implicit/explicit pressure on present activity [195]. An entity's existence and "being" can be considered to be over time and time is finite: ultimately ending with our death [166]. As a result, for both biological and artificial mortal artefacts, understanding what an authentic life is means coming to terms with one's finitude. This notion of death — and its effect on an entity's life — helps to philosophically frame mortal computation. Although philosophical treatments of existentialism and life/death [126, 166, 125] have focused on human beings and spoke to broad notions, we can adapt a small part of it for our purposes: *extracting meaning out of — and making sense of — death shapes an entity's life*. This allows one to understand that animals, including humans, with the knowledge, whether implicit or explicit, that their existence is finite, are motivated to replicate and take actions in the service of enduring persistence.

---

[1] Transformers, at the time of the writing of this article, have become far larger than those studied in [322], resulting in far larger carbon footprints. However, even then, training a BERT-style transformer [102] had the carbon footprint of a round-trip flight between New York and San Francisco, for a single passenger.

To extend the above line of thought, naturalist philosophy of mind [180, 181, 331, 342] characterizes a living entity's continual foraging of information as self-evidencing, i.e., optimizing evidence for its own existence [177], through what is known as autopoietic enactivism [97]. Underlying this enactivism are two notions: operational closure and sense making. Operational closure means that a system must undergo autopoietic self-assembly and self-maintenance to keep separate its internal states from the external states of its niche. This creates the boundary between what the system is from what it is not. Crucially, the system's (e.g., metabolic) processes determine the nature of this boundary, allowing a cell, for instance, to emerge from primordial, molecular-chemical soup [339, 341]. As a direct consequence, a system's existence can be viewed as a process of boundary conservation; this motivates our Markov blanket formalization of mortal computing in Section 4. Living entities exist by individuating themselves from everything else [331]; the processes that do this are inherently those of identity constitution, yielding a single, functionally coherent unit [341] or self-evidencing whole [103]. In complement to operational closure is sense making; namely, the acknowledgement that there is a mutual dependence between external processes and the entity's internal processes, i.e., a biological system must distinguish itself from its niche yet still be coupled to it in order to persist in that niche [341]. Sense making refers to the system's possession of operationally closed mechanisms that help infer the consequences of its sensorimotor actions and distinguish different probabilistic implications of otherwise equally viable paths through a niche [103]. This autopoietic, enactive framing of living systems will serve the biophysical, cybernetical, and cognitive framings of mortal computing in the following section.

# 3 The Biophysical, Cybernetical, and Cognitive Framings of Mortal Computation

To build a foundation for mortal computation — that resonates with a more biomimetic form of machine intelligence — this section synthesizes core concepts from physiology, biophysics, cybernetics, cognitive science, and naturalist philosophy of mind. Woven together, these ideas form the fabric for a theoretical construct of mortal computation, motivating its definition in Section 5.

## 3.1 The Biophysics of a Mortal Computer

Made possible through the flow of matter and energy – and built on autocatalytic closure [193, 145, 179] — a living system can self-organize and grow, maintaining and repairing itself as it is subjected to environmental perturbations. This is fueled by an organism's metabolism; its set of life-sustaining chemical reactions that use energy to transform materials into structures that themselves harness further energy to transform/transport material as well as eliminate/excrete surplus material and toxins [249, 266, 292, 241]. Powered by enzyme catalysts, through the metabolic pathways [212] inherent to catabolism — which destroys material to release energy — and inherent to anabolism — which synthesizes material by consuming energy — an organism is able to determine which material is toxic or worthy of consumption, reacting to changes in its environment and the behavior of other entities.

Crucially, any organism's persistence depends on the careful regulation of the above metabolic processes [350, 297, 61, 316], such that the values of key metabolic variables, such as temperature, pH, bodily fluid balance, are kept relatively constant, i.e., within certain boundaries and near a target value called a *set point*. This is the purpose of homeostasis [50] – to ensure that the system's essential variables are stable as it adjusts to an evolving, adversarial environment [71, 72, 50, 53].[2] Although the feedback-based error correction ability of homeostasis goes far to provide a biological system with internal stability, it is higher-order regulation and error prevention afforded by allostasis [320], that allows it to act in pre-emptive, non-reactive ways to ensure that homeostatic regulation is sufficient. Acting as a sort of "predictive homeostasis", allostasis provides anticipatory measures that are triggered by indirect cues beyond the reach of homeostatic control, such as stressors and surprising environmental conditions [235, 65]. Ultimately, the purpose of allostatic control is to preclude future deviations in essential variables from happening in the first place. In a mortal computer, this means that allostatic processes finesse homeostasis (or more generally homeorhesis), working to satisfy constraints before the need arises; this allows the mortal computer to budget resources, as living systems do [319, 320]. Taken together, in both living entities and mortal computers, reactive homeorhetic maintenance and proactive allostatic control work jointly to ensure that life-critical variables remain within viable ranges.

---

[2] In an entity as simple as a cell, homeostatic regulation is responsible for keeping a wide range of biophysical quantities including extracellular fluid, temperature, acidity/basicity (pH), concentration of dissolved particles (osmolality) such as sodium, potassium, glucose, carbon dioxide, and oxygen, relatively constant and near a desired value.

As a consequence of the biophysical formulation above, a mortal computer's self-organization (of its primitives/internal states related to its hardware) constitutes its thermodynamic and metabolic efficiency, grounding its agency and ability to adapt. Importantly, its metabolic organization stands far from thermodynamic equilibrium given that energy and matter would be lost as heat and waste and it thus must forage, much as in living systems as they continually acquire new resources. It is this relationship between an open system and its environment that underwrites the system's design; so as to ensure its thermodynamic costs are reduced to those of living systems. Given that the second law of thermodynamics dictates that within any closed system the amount of entropy cannot decrease, it follows that a 'living' (i.e., open) mortal computer is not in thermodynamic equilibrium and thus operates as a dissipative system. This means that it must increase the thermodynamic entropy of its environment [344, 346]). Therefore, the mortal computer's persistence depends on its ability to remain far from thermodynamic equilibrium (i.e., death) through its low-level processing (i.e., the acts taken to minimize its free energy; see Section 4) so as to establish order through generating disorder [99].[3]

The final feature — needed to fully characterize a mortal computer is that of autopoiesis — is its ability to make itself from within. For any entity to be considered autopoietic, it must continuously (re)produce, organize, and maintain itself, including its constituent parts/processes and the pattern of relationships between them, without external intervention [339, 234, 340, 60]. This means that a mortal computer must maintain its form, integrity, and function, emphasizing a morphic treatment of intelligent, adaptive behavior. Furthermore, autopoiesis, in the sense used here, encompasses homeostatic and allostatic control, given that an autopoietic system can leverage these regulatory mechanisms to maintain its internal milieu. Autopoiesis grounds a system, including its production network of processes/elements that transform/destroy, in physical space, which means that an autopoietic entity is specified and shaped by its topological domain.

Framing a mortal computer as autopoietic firmly grounds it in its physical medium and dynamic properties. This means that it must be able to persist longer than its components, in virtue of its regeneration/replication abilities. For this to be possible, certain aspects of embedding space [339] must be in place [120, 117]: **1)** base-level materials such as the concentration of substrates, **2)** an organization which embodies/encodes information, and **3)** processes or dynamics of continual revival and internal generation. These criteria have several implications for structural and behavioral details of mortal computing systems. First, such systems must be able to select correct materials, e.g., molecules, from an environment as well as contain them in the right internal concentration/balance, using homeostatic/allostatic regulation. As a result, any such system must possess a boundary: for instance, as in a cell, a permeable membrane equipped with material filtering sub-systems embedded within. Due to the dissipation induced by fluctuation-dissipation theorems, the protective body of a mortal system will itself break down as it is buffeted by environmental perturbations [354] and will thus require ongoing repair. This repair can be effected by replacing damaged/decayed components, e.g., molecular structures, and such components can be produced using raw material(s) taken from the environment through a synthesis process, which itself may break down/decay.[4]

**Biophysical Framing.** The physiological considerations above imply that life has a metabolic aspect [59] and thus the construction of a mortal computing system (Figure 2) should start from base-level functions or variables that constitute its internal states, i.e., essential variables or 'primitives'.[5] Since the mortal computer is autopoietic, these primitives and their physical structures, which are modulated and coordinated by many levels of regulation, should evince continual self-repair. In essence, the design of a complete mortal computing entity should, at minimum, consist of: **1)** a boundary around the system's internal states (the Markov blanket; see Section 4), **2)** a synthesis engine for producing the system from within, and **3)** one or more transducer-actuator subsystems[6] that enable controlled interactions with an environment, e.g., trans-boundary pumps and filters.

---

[3] In essence, this is an application of the first law of biological thermodynamics [36].

[4] Recursively, this means that synthesis processes themselves must manufacture and remake their own components as they operate, in the correct arrangement, in order to form/maintain their own structure.

[5] The exact specification of the essential variables/primitives would, in principle, depend on the exact *in silico* instantiation of a mortal computer system. In designing biomimetic entities, it might prove useful to relate or liken many biophysical variables in terms of their analogues on target hardware platforms.

[6] Transducers convert energy from one form to another, e.g., converting mechanical energy into electrical signals.

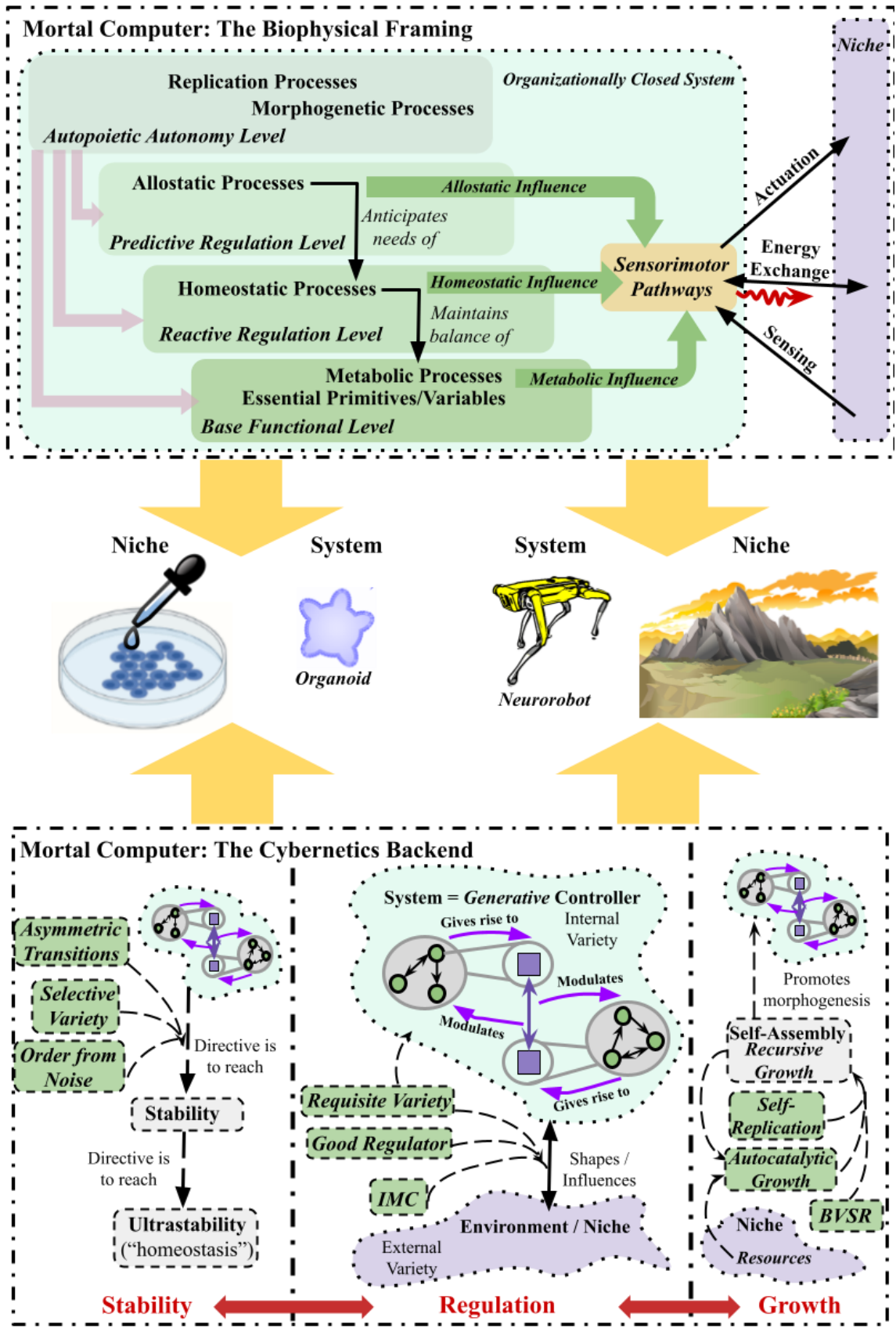

**Figure 2:** Mortal computing, the underpinnings of biological (organoid) and artificial systems (neurorobot), viewed through biophysical and cybernetic lens. A green dashed text box indicates a cybernetical principle.

A basic mortal computing unit includes a physical boundary composed of embedded transducers and actuators, allowing it to interact with its niche as an open system where material/energy pass into and from the unit across this boundary. This transduced flow is subject to the internal control imposed by homeostatic set-points: requiring simple mechanisms that compare sensed values produced by the transducers with stored, internally established set points. This production-reaction specification gives the system *personal cause for its actions*. Any comparator, in reaction to transduced signals, emits control signals to drive actuators, such as a secretion mechanism for removing waste. These transducer-comparator-actuator subsystems facilitate a mortal computer's selective interaction with its environment; variation is information, while the effect of the system is physical force

In Figure 2 (Top), we frame mortal computation in terms of conceptual levels that depict the operational nature of the requisite processes. The essential variables/states sit at the mortal computer's base functional level and relate to properties inherent to the hardware's morphology (power level/store, temperature/heat level). Above this are processes that mediate homeostatic control, i.e., the reactive regulation level, providing the measures needed to counter deviations in certain system primitives to ensure that those variables remain near characteristic values.[7] For example, a homeostatic function would trigger an internal action to route power from energy reserves to the system's primary battery, if it is below a certain level. In neuromorphic constructs of spiking neurons, one concrete homeostatic variable is the firing rate of cells, which affects system efficiency as sparse activities promote reduced global calculation and hardware resource use. This can be modeled in the form of set points implemented as adaptive thresholds/bounds that increase each time a cell fires; preventing any cell from firing too readily. At this level of homeostatic characterization, it may be useful to consider coupling homeostatic functions together: Figure 3 (Left) shows one homeostatic regulator, conditioned on transduced signals, that adjusts the set-point of another homeostatic process. A possible design choice (see Figure 3 Left) could include a stacked form of recursive manipulation [55, 119, 116], giving rise to an action-perception system or 'open-ended metabolism' [117]: here, a base transducer-comparator-actuator unit is placed within another similar base unit, but the control signal output of its set-point is wired to modify the set-point of the original base unit. Inspired by the layered, intricate patterns of the metabolic regulation observed in biological entities, where homeostasis is carried out via enzyme activity that manages the flux of metabolic pathways [316], a mortal computer would implement multiple variants of homeostatic control mechanisms. Intrinsic control mechanisms could change the operating conditions of certain metabolic functions based on higher-level set-point values, while extrinsic processes could trigger routines that copy/delete particular variable constructs, e.g., routines akin to cell division, specialization, or apoptosis (programmed cell death) [90], based on specific transduced values from outside the system's boundary.

Supervening on the homeostatic-reactive level is the predictive regulation level entailing allostatic processes. Such allostatic functions could be implemented as neural circuits, e.g., leaky integrators within a neuromorphic chip, which trigger higher-level motor command actions. These types of functions could, for example, interact with the homeostatic ones by modifying their set-points in response to information received by distal transducers/sensors within the system; see Figure 3 (Right). Allostatic functions could be made progressively more complex by integrating memory structures that recurrently capture long-term correlations in sensory streams and promote a temporal, sequential setting of set-points of homeostatic comparators. Given that a mortal computer is an open system operating at a nonequilibrium steady-state, its allostatic processes are built under the premise that homeostatic set points change due to its environment and must ensure stability in the face of such variation [318]), i.e., instantiating homeorhesis. Such higher-order processes would initiate complex pathways for adaptation in the face of a challenge, such as an external threat that would damage the mortal computer's morphology, and would shut off once the "challenge" has passed, as in living systems [235]. The top-level view, the autopoietic autonomy level, of a mortal computer includes its morphogenetic and replication processes. It is these that make a mortal computer 'autopoietic'. This layer includes descriptions of the regenerative and repairing behaviors of the levels below as well as separate creation/destruction routines, i.e., complementary neural circuit manufacturing, assignment/specialization/re-wiring routines that trigger based on a surplus of component parts, or even simple variable cloning/deletion. The specification of the generation/repairing behavior could relate to catabolic/anabolic functions of essential primitives, since such mechanisms provide the foundational production and removal engines that engender a mortal computing system's continual remaking. In addition, specification of the forms of higher-order replication would go into this description level, noting that, if mutation is introduced, mechanisms for system-level reproduction are possible. Essentially, through replication and morphogenetic processes, the system is able to self-sustain its 'identity', continuously producing itself, without external intervention.

---

[7] Given that homeostatic regulation naturally varies across organisms, determining what are the right homeostatic variables to regulate will likely be dependent on the hardware at play that will carry out core computations.

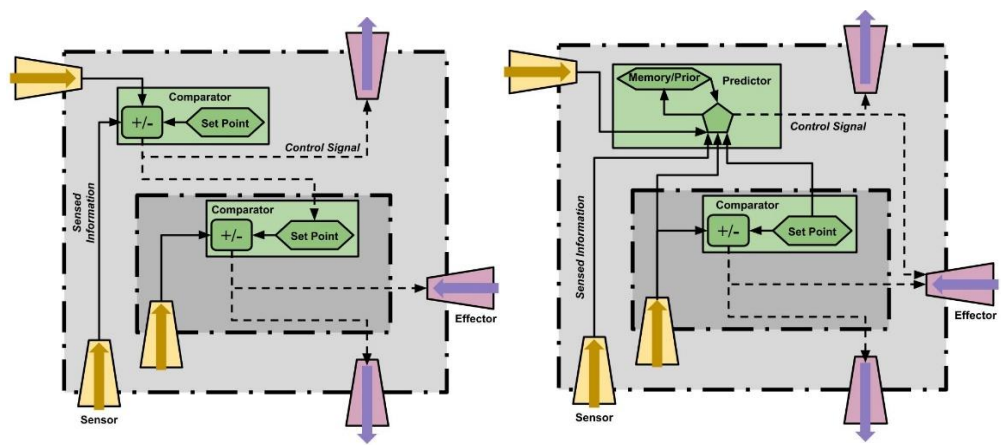

**Figure 3:** (Left) A simple mortal computer built with coupled homeostatic set-point comparators – in this system, one set-point comparator sets the set-point of another (lower-level) one. (Right) An allostatic mechanistic system – here, a predictor (that is memory-driven) anticipates the needs of a lower-level comparator, biasing the system to act so as to improve resources available to the homeostatic process. Orange modules represent sensors (transducers), and pink modules represent actuators (that control intake and secretion, depending on the direction of the flow arrow). Solid lines represent transduced information signals while dashed lines represent control signals.

Another foundational aspect of the mortal computer in Figure 2 (Top) is its sensorimotor dependencies. This is motivated by metabolic-dependent/based processing and adaptivity in biology [111]: the central impetus behind a cell's motility, e.g., chemotaxis [349, 112] or phototaxis [187, 32], and its interaction with its niche is its metabolic drive [88, 324, 12]. Sensors/transducers and effectors/actuators exhibit, to varying degrees depending on the design of the system and the morphology afforded by it, dependencies on metabolic functions and regulatory processes. In Figure 2, these are labeled as 'allostatic influence', 'homeostatic influence', and 'metabolic influence'. For instance, the operation of a motor subsystem might be conditioned on a subset of essential variables, the set-points of several homeostatic comparators, and control signals emitted from one or more allostatic controllers. How such dependencies modulate or alter the functionality of a particular sensory or actuator subsystem would depend on the system's substrate as well as its niche. It is through the sensorimotor pathway that energy/matter is exchanged with the mortal computer as it engages with its niche.

Finally, notice in Figure 2 (Top) that the dotted bounding box that encloses the mortal computer's functional levels and sensorimotor pathways is labeled as 'organizationally closed' – a mortal computer should operate solely on the basis of its own self-reproduced structures, rather than on the inputs it directly receives. While it is important for a mortal computer to be organizationally closed, this does not commit a designer to a specific physical manifestation. Simply put, organization refers to the relationships between the system's processes and parts. The mortal computer's structure refers to the physical instantiation of the constituent components and their relationships. One organization can be manifested in many structural forms. This means that, since it is autopoietic, a mortal computer's structure might change with time, but its organization persists. A mortal computer's responses are dependent on — and determined by — its structure but also triggered by environmental events. This means a mortal computer, like a living system, is constantly interacting with and thus coupled to its world as indicated by the directed arrows of 'sensing', 'actuation', and 'energy exchange' in Figure 2 (Top). In short, the full story of a mortal computer includes its niche: the mortal computer and its environment mutually perturb each other and trigger changes in the state of one another.

## 3.2 The Cybernetics Backend of a Mortal Computer

Some of the key ideas that underlie this view of mortal computing come from classical cybernetics [351, 25, 364, 39, 105, 167, 270, 169], a field that can be characterized by the core idea that any system is, in effect, a generative model that acts as a model of the environment that it controls. The structure of an organism, in and of itself, is taken to be part of a (control as) inference process. A key biophysical term that was notably associated with this framing was homeostasis, demonstrated by Ashby's famous homeostat [24]. These cybernetical notions echo many of those in physiology, strengthening the view that mortal computation must first be grounded in its structure and implicit mechanisms. At the core of cybernetics is the notion of 'retroaction' – a system's directive is to, in a somewhat Kantian fashion [192], incorporate its "ends" (i.e., goals) into the "means" (i.e., mechanisms), to ensure that attainment of its goals is (almost) inevitable. A central question that retroaction prompts is: *how is it possible for a system to learn what it needs to know, so as to act effectively?*

This question motivates a complementary characterization of mortal computation, i.e., its 'cybernetics backend' as seen in Figure 2 (Bottom), a particular synthesis of cybernetical principles or axioms [25, 167]. Foundationally, cybernetics is about the system, which is treated as consisting of elementary parts and their local interactions, which exhibit collective dynamics [178, 311] from which overall order might spontaneously arise, given sufficient energy [28, 101, 191, 22, 193, 168, 94, 352]. In such a system, the parts give rise to the 'whole' (upward causation), and constrain it, to a degree, while the whole constrains the constituent parts, making them act so as to conform to the laws of the higher levels defined by the system (downward causation) [70]. To completely understand a system in cybernetical terms, particularly its relationship to an environment, one must start with the primary currency of cybernetics; namely, variety.

Variety, simply put, is the measure of the number of distinguishable states in a system's state space [23] or $\log_2(|S|)$, where $|S|$ indicates the cardinality of the state space $S$.[8] Variety can further be interpreted as the degrees of freedom that a system has in choosing in what specific state it will be. Changes in variety characterize a system's evolution and, importantly, variety may be reduced through selection [169], which is the criteria used to classify which states are stable and which are not. To furnish a perspective on mortal computation, we partition the cybernetic principles into three complementary groups: 1) stability, 2) growth, and 3) regulation/control. In Figure 2 (Bottom), we depict the core relationships among these groups, centering around regulation/control.

**The Quest for Stability.** The primary imperative for a system, e.g., a mortal computer (see leftmost third of Figure 2 Bottom), is to reach states of stability (i.e., an attracting set of states). The ideal state for a system to end up is one of *ultrastability*, where the system has reached a form of self-organization that allows it to converge to an attracting state without altering its part-to-part relationships, i.e., it has developed feedback-driven step-mechanisms that alter internal variables in response to environment perturbations [364, 21]. It is here that the principle of asymmetric transition applies: it is possible for a system in an unstable (high energy) configuration to transition to a stable (low energy) one, but not the converse. This means that a system is implicitly performing a selection, leaving a region in state space with high variety to a region with lower energy. As the system rejects fewer states as it reaches more stable ones, the variety decreases as the system becomes more organized and, if one equates variety with (thermodynamic free) energy, a system must do work and thus exert variation. Therefore, if a system is stable, it does not undergo variation and can be viewed to be in a state with minimal energy; thus, the system is less likely to expend the energy needed to leave its current stable state.

In the effort to reach stable states, the principles of selective variety and order-from-noise apply: the greater the variety of the configurations/states that a system undergoes, the more likely it will retain at least one of these states (the first principle), while random perturbations that buffet the system will aid it in finding more stable states (the second principle). In other words, noise will increase a system's chances of finding more stable states, increasing its fitness in the long-run. In complement to this, the principle of blind-variation and selective retention (BVSR) [68, 69, 312] posits that configurations produced through 'blind variation', i.e., the system cannot know which states will be selected by its actions *a priori*, will likely lead it to transition to a stable state, which will less likely be eliminated compared to less stable ones. As a consequence, a system's evolution is conservative, i.e., stable configurations are retained, thus exploited, while unstable ones vanish [69].

**System Growth.** The second key quality of a cybernetical system is that it is morphological and morphogenetic, thus capable of growth and replication (see rightmost third of Figure 2 Bottom). This connects back to the framing of a mortal computer as autopoietic, which required that it had the ability to regenerate and self-repair, while preserving its coordinated-whole. From the cybernetics viewpoint, a system is morphological if it is able to offload or carry out some calculations — necessary for navigating an environment — to its bodily structure, i.e., its morphological form.[9] Morphogenesis then refers to the system-environment exchanges that shape the particular form of a system [332, 360, 334, 157]. A morphogenetic system, from a cybernetics point-of-view, is one that maintains its continuity and integrity by altering essential aspects of its organization and/or structure [49]. Note that a morphogenetic process may be triggered by changing environmental conditions. In biology, this could be the regeneration of multicellular structures in response to external damage, and can explain the emergence of patterns in both artificial and living systems, e.g., well-known examples include 'Turing patterns' [334] and 'morphogens' for cell type coordination/generation [360]. Self-replication, which we view as building on morphogenesis, is a system's behavior that results in identical/similar copies of itself or parts of itself. A well-known example in biology is cell mitosis, i.e., division. In cybernetics, aspects of the principle of replication come from work on cellular automata [209], with origins as early as the Von Neumann universal constructor [345] and the "game of life" [148, 48]. In essence, a base set of cybernetical qualifications — for a system to be self-replicating — is that it must generally have: **1)** a 'code' representation of the replicator, **2)** a mechanism to 'copy' the coded representation, and **3)** a mechanism for carrying out the construction within the replicator's environment [345]. A replication process that yields an imperfect copy of its source (e.g., a mutation is applied) becomes subject to natural selection and pursues its own quest for stability as described above; this "offspring" is another basis by which a system's organization — its identity — is propagated over time.

---

[8] If a system has 2 states, then there is only one possible difference/distinction; variety is equal to one.

[9] This has become increasingly important in the domain of robotics, resulting in what is known as morphological robotics [223, 276, 164, 210, 211]. For example, much as animals exploit the material properties of their legs' elastic muscle-tendon system to reduce the computation that their brains perform, robots with leg actuators work to exploit bio-mechanical properties of those legs to engage in locomotion and thus reduce required internal physics-based calculations [276].

If a system reaches a nonequilibrium steady state, and is self-replicating and morphogenetic, then the principle of autocatalytic growth [194] comes into play — a positive feedback loop is created such that explosive growth is possible and the system will replicate, and given finite resources, systems that are less fit will die due to competition. Autocatalytic growth, in tandem with the principle of BVSR, drives the self-assembly of a system that is distinguishable from its environment due to its morphological boundary; this is the cybernetical form of autopoiesis. This also gives rise to the principle of recursive system growth, where interacting self-stable systems, each with their own boundaries, acts as a set of related building blocks that will undergo variation and thus recombination to yield higher-order configurations; some of which will be more stable than others. Much as the base systems underwent selection to reach stability, so will the higher-order assembly and, as a result, hierarchical and heterarchical ("nearly decomposable") systems [310, 189] will emerge. [10] Given that autocatalytic growth underpins the principle of recursive growth, when a stable state assembly makes it easier for other building block systems to join, the number of its constituent building blocks will grow [178, 290].

**Regulation and Control.** The third and final aspect of mortal computing — that we seek to articulate through the lens of cybernetics — is its relationship to its environment through regulation and control. Here, three foundational notions come into play: the law of requisite variety, the good regulator theorem, and the principle of internal model control. The law of requisite variety [200, 27] states that a controller can only control a target system, e.g., an environment, if it exhibits sufficient internal variety to represent it; the system should embody the minimum number of states needed to obtain ultrastability [21]. In the context of active regulation, the mortal computer must have at its disposal a sufficient variety of actions to ensure that there will be little variety in the outcomes of its essential variables; it must be capable of handling the variability inherent to its environment. This, in effect, relates to a mortal computer's ability to maintain its internal states, i.e., its homeostasis, in response to external ones. If it is treated as a 'blocker' that prevents environmental disturbances (metabolic) variables, for it to be useful, this blocker should have as much flexibility as that which it is blocking; otherwise, it will fail to prevent perturbation to the variables that it blocks [280]. A consequence of requisite variety is that for a system to exert sufficient control over its environment, it should try to maximize its internal variety to be optimally prepared for perturbations that it might encounter; maximizing internal entropy increases its cybernetic freedom or options for action [119, 118].

In the context of this balance — between internal and external variety — emerges the good regulator theorem [91], also known as the law of requisite knowledge [167]. The good regulator theorem states that every self-regulating controller of an environment must itself contain a model of that environment. This theorem implies that the regulator, e.g., a mortal computer, becomes a model of the niche that it seeks to control and, more generally, that the system's survival requires forming, learning, and maintaining a model of its environment. This is essential for reaching ultrastability. A useful complement to the good regulator theorem is the principle of internal model control (IMC) [131], which dictates that the purpose of a controller that manages an environment is, first, to provide closed-loop stability, and second, to regulate a variable that is a function of the target's output and a reference signal. IMC asserts that, if a system preserves these two properties when certain system parameters are perturbed, then the "synthesis", or joint controller-environment system, is deemed structurally stable. The IMC principle helpfully connects a mortal computer's adaptation to its environment back to one of its central directions; achieving stability far from equilibrium.

**Cybernetical Framing.** As can be seen in Figure 2 (Bottom), the cybernetics backend of a mortal computer characterizes its collective dynamics, emergent behavior, and relationship with an environment based on the principles above. The mortal computer or 'generative controller' (depicted as a two-level system for simplicity), comprises low-level parts, which could be the base essential variables in Section 3.1, that have relationships (possibly bidirectional ones) with each other, as indicated by the black arrows. Clusters of these low-level units give rise to higher-level units (shown as purplish squares), which also have relationships to each other; lower-level units upwardly cause higher-level ones and higher-level units downwardly modulate the low-level ones.

---

[10] Note that larger assemblies, made with fewer building blocks at their particular (time-)scale, have a lower probability of arising via blind variation.

The imperative of the mortal computer is to reach stable configurations of low energy. Upon reaching more stable configurations, it is unlikely to return to previously stables ones of higher variety (the principle of asymmetric transition). Furthermore, due to the principles of order-from-noise and selective variety, random fluctuations in the environment enable it to continually find more stable configurations. The mortal computer's imperative is to reach ultrastability, also viewed as successful homeostasis. The environment is a crucial factor in a mortal computer's evolution towards stability; it holds the resources that the system needs to drive its self-generation processes and thus its self-assembly. Given that the mortal computer is autopoietic and self-replicating, due to the principles of BVSR in tandem with autocatalytic growth, it is able to recursively grow itself as it adapts to its niche. Finally, given that a mortal computer is generative, in that it brings about its environment through action — and learns about its properties via sensors (encoding this knowledge into memory) — it operates as a good regulator, working to be a model of internal control. The key is that, as a result of the self-assembly that promotes its morphogenesis, the mortal computer exhibits enough internal variety in comparison to the variety of its environment so as to ensure its continued persistence.

In terms of the cybernetical variety, we may then understand the evolution of a mortal computer over time through its connection to the physical world. The physical world itself consists of particles (in space-time) and energy flow (caused by forces across these particles), resulting in an ever-changing configuration/state of particles. This means that, at any one point in time, the environment is in a specific configuration that is different from any other. As a consequence, the differences between the universe's configurations are what a system will be concerned with, where information pertains to these differences [35].[11] The mortal computer perceives these differences in its interactions with its world, and responds by adjusting its internal states, thus maintaining organizational closure. As a result, information embodied in a mortal computer's configuration or structure self-organizes to correlate with the patterns that it perceives and responds to in its surroundings. Notably, cellular automata (or artificial life [209]) embody many of the above principles, have been shown to exhibit non-equilibrium dynamics [358], and, in light of our cybernetics backend, serve as a possible minimal model of a mortal computer based on simple transition rules.

### 3.3 A Cognitive-Philosophical Grounding for Mortal Computing: 5E Theory

Thus far, we have established an implementational view of mortal computation through biophysics and an analytic view of it through cybernetics. Building on these results, we now consider its cognitive-philosophical grounding. Concretely, we break down a mortal computer into its 'cognitive slices' to manifest what we will call a *5E Theory of Cognition*. In cognitive science, 4E cognitive theory [146, 337, 248] categorizes processing into four groups: embodied, enactive, embedded, and extended cognition. In this work, mortal computation is cast in terms of a form of 4E theory that we extend to include 'elementary cognition' or, in biology, 'basal cognition', which characterizes the behavior inherent to all living entities based on functionality that existed prior to nervous systems [219, 224, 123, 124]. This gives rise to a 5E theory of cognition, depicted in Figure 4, described below:

1. ***Elementary***: Cognition stands on fundamental functions and structures that enable acting and tracking aspects of a niche to ensure survival, i.e., finding food, avoiding danger, and to reproduce [229, 275] – this is basal cognition [219, 224], which manifests through a system's autopoiesis;

2. ***Embodied***: Cognition cannot be described in terms of abstract mental processes and representations alone, i.e., it must involve the entire body/morphology of the living system (the body serves as a constraint on the brain and/or a distributor/realizer of cognition) [18, 128] – this is the embodiment thesis [18, 9, 128, 342, 308];

3. ***Enactive***: Cognition is the set of meaningful relationships determined by an adaptive two-way exchange between the biological and phenomenological complexity of living creatures and the environments they inhabit and actively shape [291, 331, 147] – this is enactivism;

4. ***Embedded***: Cognition is not an isolated event, separated from the living entity's econiche, i.e., an entity displays layers of co-determination with physical, social, and cultural aspects of the world [329, 82, 63, 278, 38] – this is situated cognition or situatedness [293]; and

5. ***Extended***: Cognition is often offloaded into biological beings and non-biological devices to serve as functions that would be impossible/too difficult to achieve by only relying on the agent's own mental processes – this is extended mind theory [86, 356, 84, 185] (the mind extends beyond the boundaries of an individual).

---

[11] This is a more physical, grounded manner in which to interpret the cybernetical notion of variety.

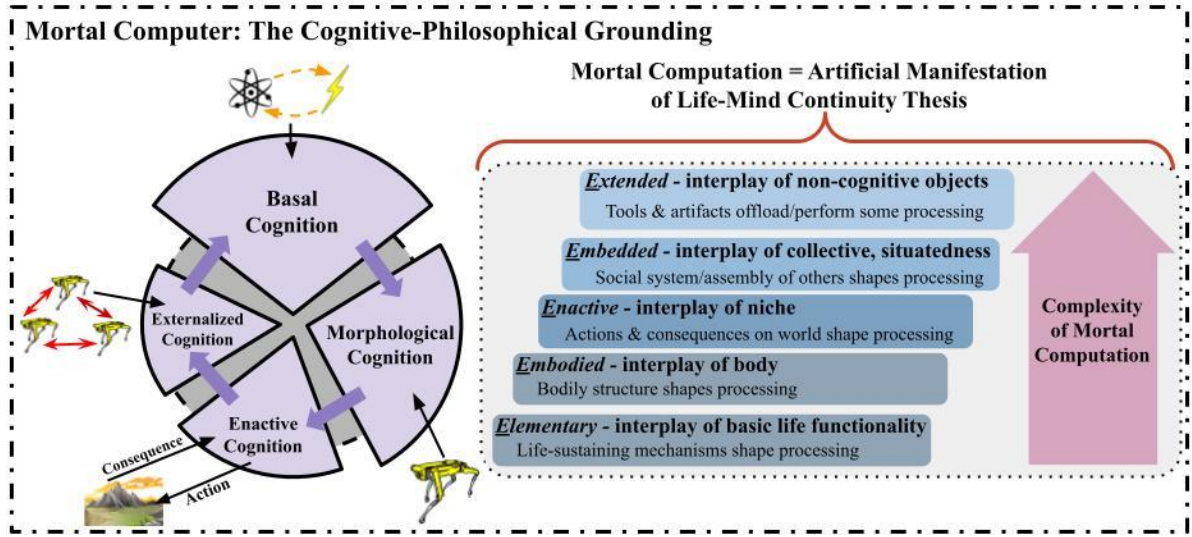

**Figure 4:** Mortal computing manifested by slices of interrelated cognition: basal cognitive functionality (autopoietic drive), morphological/bodily functionality, enactive niche-influenced functionality, and externally-driven functionality. Mortal computers, of increasing complexity, would be manifestations of the life-mind continuity thesis.

Notice that, when accounting for all five of its tenets, the above 5E theory connects not only an entity's mental processes to its life-sustaining processes [339, 234], its body or morphological structure, and its biological niche but also to the collective dynamics/processes beyond the individual system, e.g., multi-agent groups, abstract organizational constructs. This can include elements that are not physical manifestations, e.g., social constructs and cultural expectations that situate cognitive function [293], as well as non-biological objects, e.g., technological artifacts such as pen/paper or intelligent phones.

This stands in contrast to isolated cognition theories where neural/mental processes are the sole drivers of cognitive behavior [127, 160, 176, 177]. Through the lens of the first three *E*'s of 5E theory, a mortal computer can be explained as responding selectively to its niche, consistently regulating its boundaries, i.e., engaging in autopoietic enactivism [97], so as to ensure that it does not dissipate or dissolve into its environment. The success/failure of its responses to maintain its low-level primitives/states – which are linked to its biophysical homeostasis or cybernetic ultrastability – are intrinsically determined by its body/morphology [347, 15], subject to a selection history. However, while our focus has been on how mortal computation centers around how self-regulation, bodily structures, and the niche shape computations, an entity's survival and continued existence would certainly be affected by its interaction with other actors, e.g., other mortal computers, in its world as well as its use of available non-living objects/tools. Extended cognitive systems, under this notion, can then be taken to include neural, non-neural, and beyond-the-body [86] (e.g., a spider's woven web [185]) aspects of the organism's environment. It is possible to push the view of the body's role in cognitive functionality further: instead of treating non-neural structures as (auxiliary) resources that serve disembodied cognitive processes, one can view them as substitutions for what would require complex internal mental representations [308], i.e., the replacement hypothesis.

In Figure 4 (right panel), we depict the groupings of 5E cognitive theory, arranged in a bottom-to-top depiction of increasing cognitive complexity and effect. Basal cognitive functionality, which encompasses the regulatory processes that manage essential primitives as well as the base forms of selection and navigation that emerge in organisms through natural selection, is placed underneath embodied cognition; even though it manifests itself through a living system's morphology/organization. In the left half of the diagram, we depict the cognitive slices (scopes) that would characterize a mortal computer: **1)** elementary cognition (functionality through autopoiesis), **2)** morphological cognition (functionality through bodily structure/form), **3)** enactive cognition (functionality through interaction between a body and its niche and *vice versa*), and **4)** externalized cognition (functionality through the out-of-system entities including other mortal computers and non-mortal objects). The cognitive scopes, as indicated by the purple arrows, depend on and modulate each other. Morphological cognition is intimately motivated by homeorhetic/allostatic mechanisms inherent to basal cognition. Enactive cognition depends on the morphology producing actions so as to manipulate/respond to the environment thus triggering a consequence as the effect of its causal action. External cognition requires participation of the mortal computer in a collective of others, including relevant abstract norms and standards, and the results of this externalized cognitive processing would produce top-down modulating effects on the mortal computer's basal level.

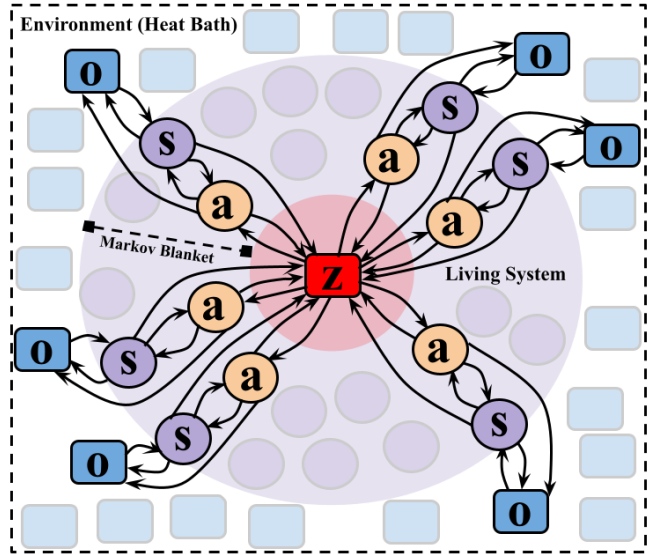

**Figure 5:** A cell in terms of its internal states $z$, environmental states o, and its Markov blanket: sensory $s$ and action $a$ states. Arcs are conditional probabilistic relationships between nodes.

Examining mortal computation through the lens of 5E theory licenses a characterization in terms of increasing complexity according to cognitive functionality. As in Figure 4, study of its behavior could be made through one cognitive slice, e.g., one might want to focus on its embodied functionality (what affordances [155] does the system's body provide) as opposed to its extended functionality (how the system uses tools). This also warrants consideration of the relationships between the cognitive scopes, e.g., functions offered at the basal cognitive base dictate the morphological cognitive functions that are possible which affect the system's enactive functionality. Note that 5E theory's cognitive scoping casts mortal computing in terms of a spectrum between simple and complex systems, much as the life-mind continuity thesis[12] does for brains/minds [144, 198].

## 4  A Free Energy Framing of Mortal Computation

Motivated by the foregoing biophysical, cybernetical, and philosophical-cognitive perspectives, we move to a theoretical backbone for mortal computation, which we will call *mortal inference-learning and selection* (MILS). MILS foregrounds the separation of the temporal scales that characterize a system and its optimal behavior in terms of three processes that all organisms engage in: inference, learning, and the selection of adaptive structures. To characterize these three processes, we start by reviewing the Markov blanket construct and ensuing free energy principle. Furthermore, we foreground the distinction between the morphic and amorphic, as this will distinguish computational systems that are inherently mortal from those that are not.

### 4.1  A Theoretical Starting Point: Free Energy

The free energy principle asserts that for entities to maintain their structural and functional integrity, they must change their relationship with their econiche through action. From a biophysics perspective, the free energy principle (FEP) characterizes the self-organization of open random dynamical systems that actively resist a natural tendency to disorder, or, in other words, remain measurable/identifiable as distinct, persistent entities over some macroscopic time-scale [137]. According to the FEP, any entity with a non-equilibrium steady-state solution (NESS)[13] possesses the following properties: **1)** its internal density dynamics are conditionally independent of the system's environment; i.e., the environment and the system are weakly coupled and the entity has states that are distinct from its external or environmental states, and **2)** it will continually 'self-evidence' [177] by returning to, or close to, its NESS; i.e., the entity behaves so as to preserve its functional integrity and its dynamics do not diverge when a perturbation is applied. Note that these two properties resonate with our earlier conception that a mortal computer is autopoietically enactive and works towards ultrastability.

---

[12] The life-mind continuity thesis states that the origins of mind are tied to the origins of systems that actively persist (living ones), as opposed to those that passively persist such as thermostats or coupled pendulum systems [198].

[13] A living system must be in a non-equilibrium state given that such a state means that there is a net flow of energy/power and/or material through it – such states are not self-sustaining and thus require implicit or explicit energy. This is a critical condition for life since energy will be used to drive the process(es) for intake/removal of matter from the system. As a result, if no energy/matter is input into the system, the system will relax to equilibrium and thus "die".

**Characterizing a Mortal Computer with Markov Blankets.** As a result of the above weak coupling and local interactions — as prescribed by FEP's two core properties — the full (joint) state space of the system, e.g., a mortal computer and environment, can be partitioned into external/environment states $\mathcal{O} = \{o_i(t)\}_{i=1}^{|\mathcal{O}|}$, internal/system states $\mathcal{Z} = \{z_i(t)\}_{i=1}^{|\mathcal{Z}|}$, and intermediate states $\mathcal{M} = \{m_i(t)\}_{i=1}^{|\mathcal{M}|}$, also known as (Markov blanket) states.[14]

The Markov blanket construct [273] acts as the interface between what is inside the entity and what exists outside of it [174] and is manifested within the morphological boundary of Sections 3.1 and 3.2. It can also be described as a statistical neighborhood that consists of all states that are parents of those in $\{z(t)\}$, children of those in $\{z(t)\}$, or parents of the children states of $\{z(t)\}$ [85]. Crucially, the Markov blanket object provides the necessary partitioning of states as required by FEP's core properties: states that are not included in the relationships described (parents, children, and parents of children of $z(t)$) provide no further information about internal states $z(t)$, meaning that $z(t)$ is conditionally independent of environment states $o(t)$ given the presence of $m(t)$. Given that they describe how an entity perceives and acts on its niche, the Markov blanket states can be further decomposed into two sets of states: sensory states $\mathcal{S} = \{s_i(t)\}_{i=1}^{|\mathcal{S}|}$ that mediate the effect of external states on internal states, active states $\mathcal{A} = \{a_i(t)\}_{i=1}^{|\mathcal{A}|}$ that mediate the effects of internal states on external ones.

To ground the notion of a Markov blanket in the context of mortal entities, consider a neuronal cell. The Markov blanket would, in this case, take the form of the cell's surface and separate intracellular from extracellular elements. The neuron's sensed environment is dictated by the sensory structures that it deploys, which include post-synaptic receptors and channels, as well as distributed biochemical and bioelectric sensors and subsystems for material pumping/secretion. Formally, the neuron's sensed environment is $\{s(t)\}$, the set of activities produced by its post-synaptic specializations, and its actions upon the environment are mediated by $\{a(t)\}$, the set of activities generated by its pre-synaptic specializations. Again, the sets $\{s(t)\}$ and $\{a(t)\}$ comprise the neuron's Markov blanket and its perception and action are crucially linked by the dynamics of its internal states which are constrained by the cell's internal structure, e.g., its genome, mitochondria, other organelles, the cytoskeletal network, etc., or — as we will see below — its generative (i.e., world) model.

The cell acts to regulate the bio-electrochemical gradients that it detects as state variations of $\{s(t)\}$. The cell's inputs and outputs are spatially organized by its morphology and the cell is able to detect local alterations to its morphology, e.g., strain on its cytoskeleton. The function of the cell and its output/actions depend on its dynamics, i.e., the internal computational message-passing architecture, and thus on its morphology. In essence, the cell's morphology determines its function, which determines the actions it takes in its environment (here the neuronal network in which it is embedded). In Figure 5, we depict a Markov blanket characterization of our neuronal cell. Note that, under the Markov blanket formalism, we may say that the cell engages in its autopoiesis by maintaining its Markov blanket boundary, i.e., cell membrane, whereas non-living systems that do exhibit persistent local dynamics do not, i.e., a campfire dissipates rapidly in the flux of the universe, as it is extinguished by a downpour of rain.

**Variational Free Energy.** Formally, the FEP claims that the internal and active (i.e., autonomous) states of an entity evolve so as to minimize its variational free energy (VFE), which is an upper bound on self information or surprisal[15]. VFE, in effect, quantifies the discrepancy between a variational density parameterized by internal states and the probability density over external states conditioned on the blanket states. This interpretation of autonomous (internal and active) dynamics follows in a straightforward way from the conditional dependencies implied by the Markov blanket partition, where an entity is defined as possessing a Markov blanket.

The functional form of VFE can be written in the following manner (see [85, 137] for details):

$$\mathcal{F}\big(m(t), z(t)\big) = \mathrm{KL}(q_z(o(t)) || p(o(t))) - \mathbb{E}_{q_z}\big[\log p\big(m(t) \big| o(t)\big)\big] \tag{1}$$

$$= \mathrm{KL}(q_z(o(t)) || p(o(t)|m(t))) - \log p\big(m(t)\big) \geq -\log p\big(m(t)\big) \tag{2}$$

---

[14] Here, $t$ refers to time and $|\mathcal{O}|$ denotes cardinality of set $\mathcal{O}$. Note that $i$ indexes a specific value of the state; omission of this index refers to an arbitrary/generic state.

[15] The improbability of an event or how unlikely an observation would be when associated with a system sensory state.

where KL($\circ||\circ$) denotes the Kullback-Leibler (KL) divergence, which is a non-negative quantity that scores the difference between two probability distributions. We further indicate that the entity has a morphology $Q$ that structures internal states. Here internal states could correspond to synaptic activity that changes quickly and synaptic efficacies $\mu$, in a neural circuit, which change slowly. The second equality shows that VFE is an upper bound on the negative log probability of blanket states; namely the self-information or surprisal of blanket states. From a statistical (Bayesian model selection) perspective, minimizing variational free energy implicitly maximizes model evidence or marginal likelihood, i.e., the likelihood of an exchange with the environment, having marginalized over external states. This reading of the VFE licenses the notion of self-evidencing; namely, a description of entities, whose autonomous states increase model evidence (to within a bound approximation).

Equation 2 indicates that an entity can be read as inferring the states of the environment, via $q_z\big(o(t)\big)$, which can be contrasted with the true state of the environment conditioned on the Markov blanket $p(o(t)|m(t))$. This contrast is, effectively, the system's prediction error, under the entity's generative model $Q$ [227, 96, 172, 365]. In other words, one can interpret the dynamics or gradient flows of autonomous states as minimizing some prediction error, based upon the predictions of sensory states afforded by internal states, whose dynamics evolve under some internal structure $Q$. Technically, the generative model entailed by $Q$ is expressed as a joint probability density over the joint states of the system.

This joint density or generative model just is the system's NESS. In other words, the NESS density furnishes a probabilistic specification of the relationship between environment/external states and the states that constitute the entity in question. Given that the Markov blanket encompasses both sensory and active states, gradient flows on VFE (e.g., neuronal dynamics) can be understood in terms of active inference [140]: namely, internal states can be read as engaged in Bayesian belief updating, while active states mediate an active exploration of the environment in service of precluding surprising (i.e., uncharacteristic) states and thus remaining 'alive'.

The FEP offers a generic theory of self-organization via self-evidencing for entities with sufficient dynamical stability to be identified over time (i.e., technically, for entities that possess a pullback attractor or set of attracting or characteristic states). Thus, an entity is considered to be a 'thing' that is distinct from its surrounding environment given that its Markov blanket assures conditional independence between its internal states and the environment/external states, i.e., its operational closure — constraining the interaction (or information exchange) between them to be local. Notice that the form of self-organization dictated by Equation 2 ties in with our cybernetics back-end of mortal computing; particularly, the good regulator theorem and IMC.

The VFE in Equation 2 further shows that this NESS density can be factorized into a likelihood for external states given the Markov blanket (states) and a prior density (over external states). It is the priors that characterize its preferred states, i.e., the attractors of cybernetical stability, and thereby underwrite internal message-passing or dynamics. Given that a mortal computer will exhibit a preference over states[16] via its NESS density, we can say that such an entity will return to the same neighborhood of states in its state-space throughout the course of its life-span. In other words, it will avoid a large number of (surprisingly low probability) states and occupy, with a high probability, a small number of states [285]. In other words, to exist as an entity, a mortal computer simply occupies the kinds of states that are characteristic of the kind of thing it is. These are the low variety states of cybernetic [ultra]stability.

This — almost tautological — expression of the FEP becomes useful when instantiating mortal computers. This follows because a mortal computer can learn its generative model (via a process of structure learning or Bayesian model comparison – see below). When a mortal computer is equipped with a generative model its autonomous dynamics, i.e., belief updating and action, are fully determined by gradient flows on the variational free energy functional of that model. Crucially, this means that one can simulate mortal computation by simply solving the equations of motion above for any given generative model (i.e., probabilistic specification of characteristic states).

Note that, with the Markov blanket considered as part of the system, Equation 2 implies that every mortal computing system is, in a sense, autonomous and interacts with only one other system, i.e., its environment. This also suggests that both the mortal computer and its environment maintain their own well-defined, conditionally independent states, where each adapts to the other's actions [205] as resources allow. This can be read as the enactive cognition of the mortal computer, as per Section 3.3.

---

[16] A 'prior preference' or prior expectations about how environment's states will evolve/unfold across time [140].

The Markov blanket is the interface between the system and environment and, as in [122, 121], can be treated as a holographic screen in quantum formulations of the FEP. In both classical and quantum information theoretic formulations, the entity does not have access to its environment's structure; it can only sense aspects of it through the mutually correlated system-environment components of the Markov blanket. The observed sector of the Markov blanket serves as the read/write exchange between the system's internal dynamics and the environment and *vice versa*. We note that the notion of a Markov blanket can be applied recursively as well, i.e., we can consider, for example, complex systems to comprise a multiplicity of (nested) blankets that engage in self-assembly [198].

**Inference.** Given the above account of VFE, the process of inference, or the performing of short-term computations, can be expressed in accordance with Equation 2, since autonomous states follow a gradient flow on variational free energy [141]. This allows us to characterize the system's information processing as well as how its internal states change with time:

$$\frac{\partial z(t)}{\partial t} = \frac{\partial \mathcal{F}\big(m(t), z(t)\big)}{\partial z(t)}.$$

This ordinary differential equation depends on the system's morphology $\mathcal{Q}$ as well as the current value of the blanket states at time $t$. Computationally, let us consider a predictive coding circuit [288, 133, 259] or Kalman filter [190, 289] as specific examples that follow VFE gradient flows; such implementations work to adjust or 'error-correct' their $z(t)$, which corresponds to the stateful activity values of their internal units or neurons (arranged in a hierarchy) as sensory data patterns are assimilated. These models adapt their dynamics, i.e., those that govern internal activity, leveraging synaptic feedback pathways to predict incoming sensory stimuli. This happens at a fast time-scale and generally occurs each time sensory states are perturbed, or data are made available. Exactly the same gradient flows play out at a slower timescale, when considering that some internal states encode external variables that change slowly (e.g., laws, contingencies and regularities in the environment). This leads to the first separation of timescales; moving from inference about states to learning model parameters.

**Learning.** Learning in a mortal computing system — e.g., synaptic plasticity in the presence of pre- and post-synaptic fluctuations — generally occurs at a slower time-scale than inference. Crucially, learning operates in accordance with Equation 2, and is responsible for storing useful information over longer periods, serving as a building block for (associative) memory that works in service of influencing the system's fast inference dynamics. The learning process of a system could be characterized as a mechanism that manipulates or alters a subset of internal variables — namely, parameters $\mu$ — that parameterize the variational density $q_z$. Specifically, this is done using the relevant gradients of VFE resulting in the differential equation:

$$\frac{\partial \mu}{\partial t} = \frac{\partial \mathcal{F}\big(m(t), z(t)\big)}{\partial \mu}.$$

This is the kind of learning found in backpropagation of errors (backprop) with an important twist: under the free energy principle, the VFE gradients can be computed locally, leading to a local energy-based schemes that grandfather extensions of predictive coding and Bayesian filtering to *learn* parameters of a generative model, in addition to *inferring* states that are *hidden* behind the Markov blanket, i.e., hidden, external or latent states.

Notice that learning depends on internal states and blanket states taking on particular values, meaning that the learning must be scheduled according to the pace of inference. Changes to parameters — e.g., a weighted addition of a Hebbian-like update to a synapse — proceed at a slower timescale, accumulating statistical regularities and assimilating causal dependencies in the environment. Using our earlier example of predictive coding [288, 259, 298], this can be achieved — in an approximate fashion — by appealing to the expectation-maximization framework [100], where inference is carried out over $K$ steps and synaptic parameters are adjusted after each set of $K$ steps.

**Selection and Structure Learning.** From the perspective of the FEP, optimization of structure and morphology means that the mortal computer will minimize its VFE or, equivalently, maximize its morphological (model) evidence. We can express this as a gradient flow based on the morphological constructs:

$$\frac{\partial Q}{\partial t} = \frac{\partial \mathcal{F}(m(t), z(t))}{\partial Q}$$

where the structure $Q$ is entailed by model parameters. In machine intelligence research, which generally focuses on providing accounts of inference (e.g., a feedforward pass in a deep neural network) and learning (e.g., adjusting connection weights for parameters via backprop [221, 294]), this type of optimization has been considered under various names: architecture search [222, 256, 113], hyperparameter optimization [47, 315, 306], structure learning [326, 328], automatic relevance determination and so on — all driven by notions of conditional computation and dynamically structured prediction schemes [42, 80, 41, 182, 162]. In statistics and probability, this form of structure learning has been cast as *Bayesian model selection* [348, 142] or 'sparsification' in the context of factor graph learning [78, 67, 149]. The aim is to develop an optimal computer by pruning model parameters to underwrite generalization ability. When expressed as a gradient descent (as above), one is generally led to schemes that shrink various parameters towards zero and eliminate them when, and only when, the VFE decreases, or model evidence increases. Neurobiologically, this corresponds to synaptic selection and pruning; of the sort one might see during synaptic homeostasis during neurodevelopment and — at a faster timescale — during the sleep-wake cycle.

**The Morphic-Amorphic Distinction.** As discussed towards the end of Section 3.1, gross structure was the embodiment of a mortal computer's organization. If we look at how morphology applies to biological entities, certain organisms, e.g., single-celled amoeba and paramecia [150, 129], possess a three-dimensional (3D) physical structure, also referred to as morphological 'degrees of freedom' [121], that has evolved over generations. A system's morphology imposes two core constraints: **1)** a thermodynamic cost, as described earlier, that comes from the irreversible encoding of information (writing a bit value to, or erasing it from, memory) [44]), and **2)** the coupling of information: an entity's physical instantiation strictly determines how its inputs are obtained from, as well as how its outputs are transferred to, its environment.

Based on this definition of morphology, we may clearly distinguish between what can be called 'morphic' and 'amorphic' entities. In essence, all biological systems are *morphic*. Take our example of the neuronal cell: organized within its cell membrane, it has a sensing structure composed of post-synaptic specializations that transduce neurotransmitters and effectors to transmit its own signals, as well as a series of pumps/secretory mechanisms. Networks of such cells exhibit their own dynamic morphology, made up of dendritic and axonal processes of various sizes that result in differential delays between signals [228, 218], where growing/regressing dendritic spines promotes/inhibits synapse formation and location-specific interneural communication [66, 75, 175, 295]. Moreover, long and short-range cortical connectivity is shaped by activity-dependent pruning [274, 287, 309].

Changes to its morphology affect the network's processing dynamics as well as its ability to adapt: its fast inference is affected by short-term activity-dependent plasticity, while learning depends on the synaptic connections of the functional anatomy or structure [269]. As a result of its impact on the dynamics related to learning and inference, morphology itself is treated as a computational resource in living entities; evincing a strategy that characterizes things ranging from neuronal networks in brains to plants and fungi and microbial biofilms [246, 363, 338, 56, 243].

In contrast to biological entities, despite modern-day computers also possessing morphologies[17], the software that is run on them does not enjoy this intimate relationship, as it is designed abstractly and independently of the computer executing it. This is what makes such programs immortal, and makes such programs inherently *amorphic*. For instance, let us consider the popular deep neural network (DNN). Writing down its characteristic inference and learning equations can be done independently of any realization in a three-dimensional environment, despite the fact that it is pedagogically drawn as a node-and-arrow diagram. The execution of the DNN's program, i.e., its inference and learning mechanisms, is then made possible by the levels of abstraction afforded by the computer's architecture, i.e., the compiler translates a high-level language down to routines for an assembler, which translates these down to binary commands that adhere to an instruction architecture set (ISA). Furthermore, the resources needed for the DNN, e.g., weight matrices containing its synaptic values and sensory data recordings, are then read from the computer's underlying memory hierarchy (as in Section 2) to facilitate its simulation over a period of time on a central processing unit (CPU) or graphics processing unit (GPU).

Note that, as we have argued thus far, since a mortal computer follows the same self-evidencing principles of any enduring entity, it is inherently morphic. Furthermore, because it is morphic, the FEP applies. This is important because it means the mortal computer will approach the lower (Landauer) limit on thermodynamic expenditure associated with belief updating or computation. This follows because it eludes the memory wall; i.e., the thermodynamic cost of reading and writing, to and from memory, required in von Neumann architectures and other amorphic computers. More generally — and intuitively — because the FEP applies to morphic computation, a mortal computer minimizes both thermodynamic and variational free energy. This follows because the FEP states that entities at nonequilibrium steady-state, i.e., minimizing thermodynamic free energy, can be described as Bayes-optimal computers, i.e., minimizing VFE.

---

[17] Computers possess a 3D form that maximizes heat dissipation capacity and minimizes information transmission delays.

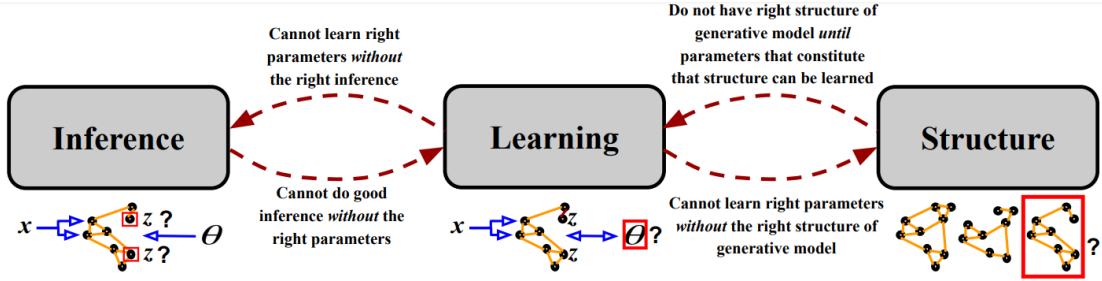

**Figure 6:** The relationship between the notions underpinning mortal computation: 1) inference, 2) learning, and 3) the selection of structure. Red boxes highlight the particular object of interest/in-question (a question mark next to the box indicates the goal or question that a particular time-scale is attempting to answer/optimize).

## 4.2 The Principle of MILS through Mean-Field Approximations

Under the FEP formulation of a mortal computer — which evinces operational closure by virtue of its Markov blanket — we saw that there was a separation of time scales, i.e., inference, learning, and selection, inherent in its 'dynamics on structure'. Specifically, a mortal computer can be distilled into the following relationships: **1)** *state inference* is just optimizing some belief structures about hidden states while, **2)** *parameter learning* is just optimizing some belief structures about hidden parameters while, **3)** *model selection* is just selecting the right knowledge of structure; in other words, optimizing beliefs about the functional form of a generative model. Nevertheless, despite being observed/occurring within their own respective time-scales, relational dependencies exist between these processes as soon as one considers the optimization perspective on minimizing VFE.

Specifically, the mortal computer cannot perform inference without the right parameters $\mu$ and it cannot learn its parameters unless it leverages the right inference (to find values for its internal states $z(t)$). Furthermore, the system cannot learn the right parameters $\mu$ unless it employs the correct structure $Q$ for the underlying generative model. Conversely, this system cannot select the right generative model structure unless it learns the right parameters. This strong circular causality (illustrated in Figure 6), which is essential in the context of adaptation to a constantly evolving niche, represents the core 'intertwinement' inherent to MILS. There is a straightforward and principled reason for this intertwinement, which follows from a ubiquitous aspect of all measurement and modelling; namely, a mean field approximation.

Applying the FEP to a mortal computer means that it is performing Bayesian inference that is *physically realized.* Given that an entity cannot physically realize exact Bayesian inference, it must resort to approximation (hence the use of a variational density in Equation 2). Variational free energy inherits the term 'variational' from the calculus of variations [151] applied to a mean-field approximation, i.e., a factorization of a probability density into conditionally independent factors [357, 362, 269]. This means that the (structure of a) mortal computer can be described in terms of mean field approximations. One might then ask: what is the typical first choice for a mean-field approximation? The answer is a mean-field approximation applied to states, parameters, and structure, which allows us to computationally separate states from parameters — and parameters from structure.[18] This means that the distinction between inference and learning rests upon a mean-field approximation and thus a conditional independence between states and parameters (and structure). We see this manifested in neuroanatomy, for example, as short-term synaptic activity versus long-term plasticity, providing neuronal networks with the ability to conduct fast inference when processing stimuli, while storing information into synaptic (associative) memory. The same conditional independence is introduced through a mean-field approximation applied to the relationship between parameters and the structure of a generative model. In the brain, a good example of this is in the way it represents the 'what'-'where'-'when' distinction when processing sensory stimuli. The physical separation of dorsal and ventral streams is often viewed as encoding 'where' and 'what' elements of visually perceived objects [336, 89]. Knowing 'what' something is does not tell one 'where' something is (and *vice versa*). The ensuing conditional independence is embodied in a morphological separation between the ventral and dorsal streams. The same separation applies to 'what' and 'when' [89, 139] given that knowing 'what' something is does not tell one 'when' it was 'there'. Ultimately, these mean-field approximations reduce complexity; thereby minimizing statistical, algorithmic, and thermodynamic costs [357]; in other words, optimizing the efficiency of the mortal computer's message-passing scheme.[19]

---

[18] Although, in some applications of classical Kalman filters [190], one could instead treat parameters as very slowly varying states – the state space is augmented with some slowly-varying variables that 'stand in' for the parameters and then the Kalman filter is just run to, in effect, 'learn' the underlying state transition matrix.

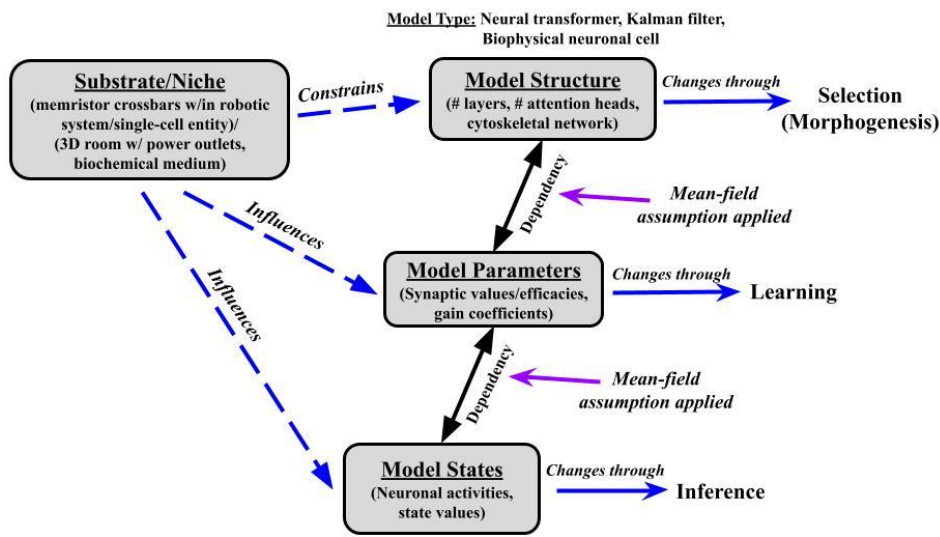

**Figure 7:** The backbone of mortal computation – mortal inference-learning-and-selection, influenced by its manifesting substrate and niche and expressly realized under particular mean-field assumptions.

The above examples imply that, in biology, it is acceptable, as well as efficient, to work under such mean-field approximations. This equally applies to mortal computers, given that they share the same goal of attaining some open (nonequilibrium) steady-state exchange with the world. However, the premise behind using variational calculus to leverage these mean-field approximations is that the system will work to 'repair' the falsehood induced by such approximations, i.e., it must compensate for the statistical independence assumptions between its states and parameters, and parameters and structure. This requires mechanisms to pass signals between the different levels, motivating a message-passing scheme that links time scales. For example, in neuronal networks, experience encoded in activity-dependent plasticity links inference to learning. A nonequilibrium steady-state entails a (VFE) gradient flow at the time scales of inference, learning, as well as the selection of structures, meaning that systems, such as mortal computers, must embody a form of activity-dependent adaptation, or, in other words, they operate as dynamics on structures with dynamics [11].

A consequence of this type of dependent adaptation is that morphology, or structural instantiation, becomes existentially important to the mortal computer, since it defines the sparsity of message-passing (of the learning and inference processes) that can occur across the system's elements. Structure becomes a core computational resource to be employed to ensure a mortal computer's continued survival and entails a search for mechanisms that do the right kind of pruning/growing to facilitate autopoiesis, e.g., choosing the right synaptic connections in a neuronal cell network in order to obtain the right graph structure.[20] In biological systems, this morphological self-organization is concerned with isolation and shielding: enzymes and catalysts, gap junctions and other controllable bio-mechanisms set up signaling pathways and boundaries [233, 361]. And this occurs at multiple scales, ranging from subcellular organelles (which partition biophysical, chemical reactions) to organ compartment boundaries to animal group behavior [307, 98, 92]. In effect, the base structural elements of a mortal computer, much as cells in multicellular structures, need to occupy their own place in a distributed morphology. This amounts to autopoietic self-assembly where each element, equipped with its own boundary, detects local patterning signals as predicted by its own generative model [141] and thereby minimizes its VFE, thus inferring its correct location and function within the overall ensemble [141, 268].

---

[19] Even in machine intelligence, such mean-field approximations apply – in DNN-based models, there is a distinct conceptual separation between the values of the synaptic parameter matrices and the number of levels/layers in a deep variational autoencoder [197] or the number of attention heads within a transformer block [343].

[20] A possible implementation of the system's morphogenesis could involve starting at a particular 'seed' (spatial) arrangement of parameters and incrementally, temporally switch things on or off, using mechanisms such as the gating variables, which could be read as a form of precision weighting [134], often employed in cognitive control architectures [258], or 'structure learning variables', that will be required to minimize their own variational free energy functionals.

The relationship between a system's morphology, its learning, and its inference underwrites a mortal computer's physically-realizable Bayesian inference, which is the engine behind its existence. According to MILS, through the FEP, morphology, learning, and inference mechanisms are all subject to the same essential mandates, each responsible for minimizing the VFE bound associated with it. As a result of Equation 2, and in accordance with the good regulator theorem, the mortal computer's morphology will need to accurately describe its environment, while also being as simple as possible [135, 136]. Finally, as a consequence of its morphology — and its coupling to the niche across its Markov blanket — the details of a mortal computer and its MILS will be reflected in the circumstances that it is instantiated under and further mediated by them. See Figure 7 for a visual depiction of the relational dependency structure that underpins a mortal computer.

## 5    The Mortal Computation Thesis

Our framing of intelligence through mortal computation is that an agent's behavioral repertoire stems from a primal impetus to preserve its identity, to remain in a non-equilibrium steady state of stability in the face of a changing environment[21]. Most of what such an agent does is in service of its basic "needs", satisfying its lowest-level requirements. Three different perspectives in Section 3, the biophysical, the cybernetical, and the philosophical-cognitive, were developed to offer implementational, analytical, and theoretical lenses through which to view mortal computing. These perspectives served as the foundation for, in Section 4, a formalization of a morphic and mortal form of computing, through the free energy principle and attending Markov blanket construction, to enable mortal learning-inference-and-selection (MILS). With the context provided by these sections in place, we now turn a definition of a mortal computer (MC) that offers an umbrella concept for implementations of mortal computation and biomimetic intelligence.

**Definition** (Informal). *a computer is mortal, if and only if:*

1. *Its morphology is specified in terms of a Markov blanket (i.e., a boundary comprising sensory and active states) and its internal structure (i.e., a generative model that entails relationships between the Markov blanket and its niche); i.e., the system is strictly morphic (not amorphic).*[22]

2. *It specifies a set of primary imperatives (i.e., characteristic states) that underwrite its persistence, and which lead to resource-oriented action (this includes designation of 'homeostatic' states, e.g., fatigue, overheating, et cetera).*

3. *It instantiates primary imperatives (i.e., precise priors over essential variables) that can be read as a generalized homeostasis (e.g., 'artificial metabolic pathways').*

4. *It implements mortal inference-learning-and-selection (MILS), with each of the requisite processes specified by a free energy functional of a generative model.*

5. *The computational processes (software) and the execution substrate (hardware) are co-designed, such that the message passing, belief updating, and generative model — underwriting MILS — can be physically realized.*

Under this definition, an instance of software has no meaning without the hardware that instantiates it; it is strictly morphic as per (Definition, property 1). The calculations carried out by a MC are grounded by morphological and thermodynamic constraints that rest on their structural realization. At its lowest level, the MC, as a consequence of its MILS (Definition, property 4) and generalized homeostatic control (Definition, property 3), will exhibit thermodynamic efficiency, sustained through material/energy pathways, as provided by its niche. While MC subsystems might not be directly influenced by global morphology, each would sense local fluctuations in its substructure and be capable of self-repair. Under the above definition, such a system would be enactive: an MC, through MILS, entails morphogenetic processes that underwrite its physical structure, encoding environmental information, given the first condition that its niche must be specified (via a generative model), thus pairing the body with an environment (Definition, property 4 and property 1). Due to property 5, which goes hand-in-hand with property 1, it follows that the MC's inference and learning processes are affected by — and are inextricably bound to — its morphological state and evolution. On one end of the causal chain (in terms of Figure 6), a hardware-software feedback loop must exist, given that the substrate dictates the set of morphogenetic functions that a MC can take. On the other end of the chain, the inference dynamics themselves vicariously inform the direction that the MC's morphogenesis will take when selecting possible structures.

---

[21] Strictly speaking, the FEP inverts this argument and states that if something exists, in the sense of possessing a Markov blanket over macroscopic time, then it acquires a teleological description in terms of self-evidencing; namely, self-organising to its characteristic states.

[22] It is clear from this that the design and specification of the system's computation strongly depends on the properties of the nature or "biology" of the hardware structures that will implement it.

Given the above, we can now organize mortal computation into three broad classes. Namely:

1. **Homeostatic MC:**[23] such a system satisfies properties 1-5 of the definition, i.e., it accounts for homeostasis over essential variables;
2. **Allostatic MC**: such a system evinces a homeorhesis that is contextualized by allostasis;
3. **Autopoietic MC**: this class evinces homeostasis and allostasis but features an autopoietic aspect; namely, self-repair and self-replication.

An autopoietic MC could serve as an ideal candidate for a complex adaptive system that instantiates a form of AGI. Such a system would be far from trivial to construct, as it would require allostatic processes — that build on top of reactive homeostatic functions — before higher-level cognitive processes could be implemented or emerge. A complementary direction for achieving this goal, as afforded by theoretical frameworks for assembling modules of a deep (i.e., hierarchical) allostatic MC, could draw inspiration from cognitive architectures [207, 321, 258].

An allostatic MC could be built with homeostatic MCs (echoing the spirit of Figure 3). Notably, to craft this kind of MC, the Markov blanket framing of Section 4 could be deployed in a recursive, hierarchical manner. As a result of their formulation, Markov blankets allow the deployment of blankets of ever-increasing scale. This facilitates the construction of organs from cells/tissues, bodies/individuals from organs, etc., where the statistical form/partitioning applies from the macroscopic down to the microscopic level. In this setup, a Markov blanket — and its internal states — at the macro-level, are constituted by the blanket states of smaller entities at microscopic scales. This hierarchical construction would then result in the dynamics of the states at one scale exerting downward casual mediation/constraint on Markov blankets in the level below [198], thus ensuring that the system minimizes its joint VFE, globally. In addition, one could appeal to principles in complexity theory [178, 290] and synergetics [161] to develop schemes for self-assembly, designing a process where Markov blankets are involved in re-configuring specific dependencies between internal, external, active, and sensory states. This self-assembling could be cast over time-scales, where slow ensemble dynamics arise from micro-scale dynamics that unfold over fast(er) time-scales: yielding an orderly coordination of the lower-level Markov blankets [290]. This implies that an autopoietic MC must comprise multiple Markov blankets and its microdynamics would be constrained by its macrodynamics. In Table 1, we review several extant mortal computers, such as the homeostat [24, 327, 74, 21, 279, 283], the electrochemical thread [271, 39, 40, 54], xenobots [30, 220, 57, 87, 56], fungal computing [245, 19, 6, 7, 16, 8], and cell preparations such as organoids [163, 314, 313] (with further details in the supplement).

## 5.2    Implications for Biomimetic Intelligence

**Directions for Inquiry.** Our theoretical framing generates several key questions that future work in biomimetic intelligence could address through the lens of the mortal computational thesis. Among the most important will be: *What can real-world, biological systems and their niches tell us about mortal computers? How might their internal functioning (e.g., design of their internal states and requisite Markov blankets) inform the design of MC systems?* Biophysical entities, e.g., xenobots [30, 220] and organoid micro-physiological systems [163, 314], offer compelling examples of systems capable of self-organization and self-replication, as well as different levels of basal cognition and collective behavior. Beyond studying these entities for patterns and mechanisms that could inform the design of *in silico* analogues, the definition of a mortal computer does not commit to implementation in any particular substrate/niche. Therefore, xenobots and organoids serve as plausible candidates for MC development. However, at the time of writing, such biophysical systems are neither readily available nor easily accessible to those that work in the sciences of the artificial, as opposed to those that work in the biological domains. This motivates work towards creating biophysical or chimeric mortal computing systems as well as exchange of insights from xenobot/organoid research to in silico system/neuromorphic engineering. In the supplement, we examine and study neuromorphic computing chips and neuromimetic schemes, e.g., predictive coding and forward-only learning [31, 254, 247], in the context of mortal computation as well as a re-framing of the machine intelligence evaluation paradigm, introducing a problem unique to mortal computation known as the body-niche problem.

---

[23] In the supplement, we sketch a homeostatic MC based on predictive coding, inspired by chemotaxis models [111].

| Homeostat | Ashby's homeostat [24, 327, 74, 21], and its subsequent updates [327], is a powerful example of a real-world adaptive regulator implemented in a non-biological medium [74, 279]. It is one of the first real-world examples of a mortal computer (MC) that implements an internal model of control that exhibits requisite variety, ultrastability, as well as self-regulation. The homeostat is able to maintain stability in the face of environmental perturbations and its imperative was to keep the value of a control variable near a target value within some specified range. The system would evaluate whether a particular set of values, i.e., a configuration (of stepper switch combinations), made it an effective controller of its environment and, if the controlled variables were not within desired ranges, the homeostat would select another configuration. Notably, the homeostat's search through configurations only took as long as the time needed for mechanical switching and the electromechanical circuit to reach its steady state, thus providing an example of computing where the only costs are those of the system's thermodynamics (Section 2). In addition, the homeostat could adapt to physical damage [283] and originally learned via a process of trial-and-error [24, 21] (making it a homeostatic MC) though later it was extended to incorporate memory-based adaptation [73] (making this version an allostatic MC, i.e., an "allostat"). |
|---|---|
| Electrochemical Thread | The electrochemical thread (or "ear") of [271, 39, 40] — also referred to as a "maverick machine" [54] — is a motivating example of what an autopoietic MC might look like. The system was made up of a few platinum electrodes inserted into a dish that were connected to an electrical source; in this setup, metallic iron threads would form between electrodes through which current would flow. These threads would exhibit low resistance relative to the solution and current would flow through them upon repeated electrical activation. If no current passed through a thread, the thread would dissolve back into its acidic solution. Over time, the threads would self-assemble into larger stable, cooperative structures by absorbing neighboring unstable sub-structures. Notably, the longer that a thread network stably grew, the slower it would break down and the faster it would return to its original structure upon a reset of the current; this demonstrated its acquisition of ultrastability. Given positive reinforcement, the electrochemical thread would respond by growing sensor organs of its own accord and incorporate them into its morphology [271], thus making it capable of creating filters conducive to its survival. |
| Xenobots | The xenobot system [30, 220, 57, 87, 56] is, in effect, a swarm of biological robotic agents that self-organizes in response to the structure of its niche. An individual xenobot is a small, self-healing biological machine based on frog cells, and when grouped into a collective, is capable of moving a payload and engaging in cooperative behavior. The more recent form of xenobots has been shown to capable of self-assembling into "spheroids" [56, 57] and individual entities within it demonstrate the ability to spontaneously specialize into new roles, resulting in new bodily behavior and designs without waiting for a slow evolutionary process to select such useful features. We remark that a xenobot system offers a concrete instance of the kind of morphogenesis that an autopoietic MC should exhibit. Specifically, a single xenobot exhibits the ability to grow, mature, self-heal, and, more importantly, contains a metabolism; it is able to absorb and break down proteins and chemicals, which makes it capable of surviving for months (in an environment containing nutrients). Xenobots not only provide insights into how basal cognition emerges in multicellular organisms but also exemplifies implementation of mortal computation in terms of a biological substrate; the "hardware" includes the genome and emergent cellular structures, while the "software" is the cellular communication that underwrites the creation of the higher-level structures. With respect to our definition's tenets, a xenobot system provides a strong representation and specification of each and could serve as an inspiration for the design of in-silico MCs. |
| Organoids | Recently, an in vitro preparation known as "intelligence-in-a-dish", or organoid intelligence [163, 314, 313], has been demonstrated to be capable of simple reinforcement learning in tasks such as playing an Atari video game like Pong [314] and has been argued to play an invaluable role for understanding biological information processing [138]. Concretely, organoids are 3D tissue cultures that are derived from stem cells and are capable of receiving genetic level instructions to drive self-organization as well as growth into complex morphologies, e.g., tissues and minimal organs. The study of "sentient organoids" [138] could not only result in new energy-efficient computational models of neural dynamics and synaptic plasticity but their natural growth and decay could foster new ideas for emulating the structural evolutionary component of an MC's MILS, which drives its ability to consolidate knowledge at the time-scales of learning and inference [236]. Given that they embody the five central tenets of our definition of mortal computation, cell cultures and organoids stand to offer a promising guide for informing the advances to be made in the design of autopoietic MCs, possibly those that will be developed at the intersection of neuromorphic-based cognition and the bio-engineering of neuronal micro-physiological systems. |
| Fungal Systems | Funguses and molds have been shown to work as computing devices [245, 19, 6, 7, 16, 8] by reprogramming their internal geometrical calculations (such as those performed by the pink oyster mushroom) and leveraging the resultant electrical activity to drive circuits. In terms of mortal computation, mold-based computational systems have been shown to demonstrate abilities related to resilience, development of intricate structures for information transmission, self-maintenance, and growth [7]. These abilities that could serve as possible ingredients in crafting a soft autopoietic MC. |

**Table 1:** Extant mortal computers – we highlight and review five real-world examples of mortal computation systems.

**Implications for AI/AGI.** There are misconceptions related to the current state of AI [3, 1, 4, 2], largely driven by our fascination with intelligence and compounded by our tendency to anthropomorphize. This is exemplified by our captivation with large language models that are intelligent in some sense but are not intelligent in the way that organisms, such as humans, are [2, 81]. The future of AI/AGI, in accordance with mortal computation, lies in the pursuit of other definitions and formulations of intelligence, primarily grounded in biomimetics and bionics. This review's aim is to illuminate a pathway that builds on many years of converging thought in biophysics, biology, cybernetics, cognitive science, and naturalist philosophy. In short, mortal computation is one possible key to AGI.

This review advocates for a different perspective on AI; current forms of AI generally treat it as a tool in support of human endeavors. Efforts in this direction eschew or dismiss other perspectives and ideas that could lead to breakthroughs in the design of general and multi-purpose intelligent systems. If we look at humans, animals, and organisms, at large, for examples of natural intelligence, we observe entities capable of performing and learning many different kinds of tasks and, if something goes wrong, such entities die. It is this very mortality that drives the mortal computation thesis and motivates a switch from a needs-oriented, 'tool-centric' view of AI (which has led to the development of powerful yet narrow AI [3] systems) towards one that grounds intelligent, adaptive behavior in the roots of life itself. The pursuit of life/survival and its relationship to mind can teach us about intelligence that is embodied, enactive, and capable of general functionality [132, 277]. The mortal computation thesis offers a possible catalyst for the next stage of AI research: Moore's law does not apply beyond a certain point and other computing paradigms, such as quantum computing [252], might provide a faster way of conducting the kinds of calculations we do today. However, this still keeps us within the tradition of immortal computing. One might view the future as residing in not viewing intelligence as a difference engine but instead as an artificial form of sentience: one that is capable of self-healing and self-repairing with autonomy — its existential imperative being to persist in (generalized) synchrony with its world.

## 6 Conclusion

The pathway forward for biomimetic and bionic intelligence offered by mortal computation, as presented in this review, synthesizes the intellectual pursuits of countless theoreticians, scientists, and engineers across the centuries. Its essence — lying at the intersection of cybernetical, biophysical, and philosophical constructs — is a groundwork for a road towards sentient behavior in that of the artificial. The ethical implications of such an artifact which, according to some perspectives, would endow it with a moral status [296, 159], are vast. If the enterprise of mortal computation is to succeed, these issues will need to be considered; ranging from the nature of the relationship of mortal computers with living entities to their rights [199, 83, 159]. If the premise of a mortal computer is correct, then the right levels of self-organizing, self-regulating processes, interacting with, and instantiated by, elements of a physical morphology capable of regeneration will engender increasing degrees of behavior ranging from basal cognition to planning and imagination. This, in principle, would possibly answer, in the affirmative, the question whether an artifact — given that it effectively is a life form — can reason, suffer, or communicate. Much as with other forms of life, such as animals, this will require consideration of the rights of *in-silico* mortal bionic entities [46], as the lines between the artificial and the natural become blurred. The benefits of such a sentient artifact, grounded in its physical manifestation and self-caused impetus for continuance, are many-fold; such an entity, as a consequence of its bionic drives and goals will exhibit a curiosity about the environment in which it exists [143, 302]. It will have the capacity to 'care' for/about others, possibly facilitating emergent empathy, social intelligence, and communicative ability. Unlike modern-day forms of computational intelligence, e.g., large neural transformer generative systems [64, 323, 77, 230, 333], that do not have an experience of body, a bionic entity might very well even be able to "...break its own rules".[24]

Despite these grand implications, this review is intended to highlight provisional steps towards 'mortal computing'. Many core ideas behind mortal computation have emerged over many decades. Mortal computing, as we have argued, may play an important role in the long-term future of research in AGI; further catalyzed by the advances in the manufacture and study of biological computers. Applications of AGI under mortal computing are manifold; ranging from edge-based computing and robotics to green AI to systems that can be deployed in the service of sustainability at a number of levels.

### Acknowledgements

We would like to thank Christopher Buckley for useful comments/pointers during the early presentations of this work as well as the Theoretical Neurobiology Group (TNB) for stimulating discussions, questions, and comments that shaped the thinking and motivations underlying this manuscript. We also would like to thank Alexander Ororbia (Sr.) for reviewing and providing edits to the early drafts of this article. This research was funded in whole, or in part, by the Wellcome Trust [203147/Z/16/Z]. We further acknowledge the support of the Cisco Research Gift Award #26224. For the purpose of Open Access, the authors have applied a CC BY public copyright license to any Author Accepted Manuscript version arising from this submission.

---

[24] Jasia Reichard, International Conference on Robotics and Automation 2023 keynote, "In Praise of Strangeness".

## Supplementary Material

### A.1 Further Questions Related to the Mortal Computation Thesis

Questions (and sub-questions) that we foresee as relevant and/or essential to the enterprise of mortal computation research are:

- What are useful measures of the thermodynamic efficiency of a given/particular mortal computer and how might this facilitate a basis for comparison/benchmarking?
- What are useful, minimal niches or environments that could be used to prototype, develop, and analyze mortal computers? Can these be simulated without requiring access to physical material and, if so, what discrepancies are introduced by relying on synthetic/simulations of a (given) niche?
- What is the base set of homeostatic maintenance functions, and how would they be designed, that a mortal computer needs in a given environment? *Notice that this question will require different answers depending on the chosen bodily substrate and target niche.*
  - How would an allostatic process be developed to automatically maintain/orchestrate these homeostatic processes and what interactions might develop?
  - How do different homeostatic/allostatic functions affect higher-level emergent cognitive ability/behavior? How does morphology affect it?
  - How much functionality should be initially designed versus how much should be allowed to (possibly) emerge? Is it possible to craft the relevant inductive biases/priors without using an outer evolutionary process?
- What degree or level of fidelity is needed to properly simulate a mortal computer in the absence of access to real-world substrate or niche(s)?
  - Does the possible domain distribution gap induced by simulating bodies/niches prevent/hinder effective mortal computing design and analysis?
- What would mortal computation look like if the internal morphology consists of neuronal elements that communicate via spikes/spike-trains? In this case, what are useful substrate-niche candidate pairings related to neuromorphic edge-computing that would foster engineering of morphogenetic, self-replicating in silico systems?
- What would a cognitive (control) architecture look like that fundamentally adheres the principles of a mortal computer that exhibits elementary, embodied, and enactive scopes of functionality?
  - What aspects of embedded and extended cognitive functionality emerge when collectives or groups of mortal computing entities interact with each other, either adversarially or cooperatively?
- What can be learned from failures and successes of biological/soft computers and/or chimeric computing systems and robots to further assist in the formulating of a mortal computer system?
  - How do degenerate states, or those pertaining to poor "health", emerge in a mortal computer? How does the MC's behavior change as a result?
- Can we design essential variables that engender/facilitate human value-aligned intelligent systems?
  - What would this look like in terms of morphological encoding/structural constraints/set-points?

### A.2 Extant Real-World Mortal Computers

**The Homeostat.** Ashby's homeostat [24, 327, 74, 21], and later updates [327], is a powerful example of a real-world adaptive regulator implemented in a non-biological medium [74, 279]. It is one of the first real-world examples of MC that implements an internal model of control and that exhibits requisite variety, ultrastability, as well as self-regulating processes. The homeostat was a control system that could maintain stability in the face of itinerant environmental perturbations. The homeostat's imperative was to keep the value of a control variable near a target value within in some specified range and the system would evaluate whether a particular set of values made it an effective controller of its environment. If the controlled variable was not in its desired range after a period of settling time, a stepping relay would choose another random set of values. The homeostat's search through the possible stepper switch combinations only took as long as the time needed for mechanical switching and the electromechanical circuit to reach its steady state, providing an example of computing where the only costs are those of the system's thermodynamics (Section 2). Furthermore, the homeostat demonstrates the cybernetic principle of BVSR – its parts (subsystems) that do the 'selecting' do not have any 'understanding' or knowledge of the detailed processes that they control or what is the optimal next action to take. Intriguingly, the homeostat could adapt to physical damage, i.e., a wire is cut, given that the stepper switch could substitute another circuit configuration in such a context [283]. Notice that the original homeostat adapted according to a trial-and-error reactive process [24, 21] and is, by this paper's definition, a homeostatic MC, embodying all five properties. An important extension of the homeostat was to include a memory mechanism that would ensure that past successful trials could bias the system's adaptation, speeding up its convergence to stable states [73]. This strengthened the homeostat system with respect to property 4 of our definition, as the learning became more sophisticated and provided the system with some measure of anticipatory mechanisms, thus making the homeostat an "allostat" and more akin an allostatic MC.

**The Electrochemical Thread Network.** The electrochemical thread (or "ear") of Pask and Beer [271, 39, 40] — a "maverick machine"[25] [54] — could be considered an intriguing example of a mortal computer, particularly one that demonstrated a form of learning, adaptation, and morphogenesis with self-replication properties. The system itself comprised a few platinum electrodes inserted into a dish of an iron sulphate solution that were connected to an electrical source; in this setup, metallic iron threads would form between electrodes where maximal lines of current would flow. Such threads would exhibit low resistance relative to the solution and, as a result, current would flow through these lines if electrical activation were repeated. Furthermore, if no current were passed through a thread, the thread would readily dissolve back into its acidic solution. Interestingly, these threads could, over time, self-assemble into larger, more stable, cooperative structures by absorbing neighboring unstable sub-structures. It was observed that, if a stable thread network grew — and the current to the electrodes was redistributed — a new network would emerge. However, if the same current were reset to its original distribution, the thread network would regrow its original structure. The longer that a thread network stably grew, the slower it would break down and the faster it would return to its original structure upon a reset of the current (demonstrating ultrastability).

The thread network's response to sound waves and other environmental perturbations, e.g., vibrations, was studied. The system was able to distinguish between sounds of different volume (50 Hertz versus 100 Hertz), magnetism alterations, and pH differences. In addition, the electrochemical thread system would respond, given positive reinforcement, by growing sensor organs of its own accord [271]. The system was able to create filters that would be conducive to its survival and incorporate them into its morphology.

**Xenobots.** A more recent development in soft robotics is the xenobot system [30, 220, 57, 87, 56], which exhibits emergent 'intelligent' behavior from a swarm of simple biological robotic agents that self-organize in response to the structure of their environment. A xenobot is a small, self-healing biological machine based on frog cells, capable of moving a payload and, when grouped into a collective, able to engage in cooperative behavior. While the first generation of xenobots were constructed via the manual (top-down) surgical shaping of frog cardiac and skin cells [57], the second generation was created from African frog Xenopus Leavis stem cells [56], which were shown to be capable of self-assembling into "spheroids". After a few days, some of these cells differentiated into cilia that served as "leg" motor actuators, which facilitated rapid movement around a surface. Promisingly, xenobot systems demonstrate that cells, removed from their original context, e.g., a frog body, can spontaneously specialize into new roles; creating new bodily behavior and designs without waiting for a slow evolutionary process to select such features.[26]

One of the most astounding properties of xenobots — that might serve mortal computation well — is that such biomachines can specialize to act as sensors, motors for locomotion, elements of message-passing networks, and mnemonic devices to facilitate information storage. This specialization is due to the xenobot's ability to grow and mature, which stands in contrast to artificial, metal-based systems: a xenobot self-heals, demonstrating a capacity for repairing and restoring itself. This offers a concrete instance of the kind of morphogenesis that an autopoietic mortal computer should exhibit. More importantly, a xenobot contains a metabolism; it is able to absorb and break down chemicals, synthesizing and excreting proteins and chemicals, and is thus capable of surviving for months (if placed in an environment containing nutrients). Xenobots not only provide insights into how basal cognition emerges in multicellular organisms but also exemplifies implementation of a mortal computer in terms of a biological substrate. Bio-machines like xenobots represent a manifestation of the co-design (property 5 of our definition) that mortal computing rests upon; the "hardware" includes the genome and emergent cellular structures, while the "software" is the cellular communication that underwrites the creation of the higher-level structures, i.e., tissues, organs, and limbs. With respect to our definitions, a xenobot system provides a strong representation and specification of each and could serve as an inspiration for the design of in-silico MCs.

---

[25] A computing engine that could emerge out of numerous components without the components' function being defined a priori. In effect, the components serve as building material for the engine's self-assembly into many possible entity structures, depending on the environmental perturbations it receives [271].

[26] It is worth pointing that the base design for the specific xenobot bodies was found via a separate, simulated evolutionary process so as to side-step the cost of a more natural evolutionary/selection process.

**Fungal Computing.** Another important example of a mortal computer is one based on fungus or molds [245, 19, 6, 7, 16, 8]. Fungal materials have been shown to work as computing devices: by reprogramming the geometrical calculations — performed internally by mycelium networks, such as those that characterize the pink oyster mushroom — the resultant electrical activity of the fungi can be used to create computing circuits. The complex internal communication that occurs via spikes inside a mushroom can be translated into messages of an efficient information transmission system that could serve as a building block for mortal computation. Notably, fungal systems demonstrate resilience, a capacity for self-maintenance, as well as rapid growth [7]. These qualities afford another promising instantiation of a simple MC; though the specification of the implicit generative model is far less clear than that of xenobot systems.

**Organoid Intelligence.** Perhaps one of the more extreme examples of mortal computation in a biological substrate is an in vitro preparation known as "intelligence-in-a-dish" or organoid intelligence [163, 314, 313]. Organoids are 3D tissue cultures that are generally derived from stem cells, capable of receiving genetic level instructions to self-organize and grow into complex morphologies, such as tissues and minimal organs. "Sentient organoids" [138] already stand to play an invaluable role for understanding biological information processing and careful study of them could result in new energy-efficient computational models of finer-grained neural dynamics and synaptic plasticity models.

Cell cultures and organoids possibly serve as a target for developing mortal computers as they are a living, growing multicellular system, and embody all five properties above. Furthermore, organoid systems have already been shown to be capable of simple reinforcement learning in tasks such as the Atari video game Pong [314]. The natural evolution of an organoid structure, i.e., its "growth" and "decay", could provide new ideas for emulating the structural evolutionary component of MILS, undergirding a mortal computer's long-term generalization ability through the consolidation of knowledge acquired on the faster time-scales of inference and learning [236]. Crafting MCs of greater complexity might benefit from the advances made at the intersection of neuromorphic-based cognition and the bio-engineering of neuronal micro-physiological systems.

## A.3    On Neuromorphic Systems and the Body-Niche Problem

**Neuromorphic Systems of Biophysical Neurons.** A neuromorphic computing chip — and other promising hardware/substrate candidates, such as liquid crystal spatial light modulators [76, 109] and Field Programmable Gate Arrays (FPGAs) [206, 253] — offers a useful (low-energy) pathway for mortal computing and resource-constrained edge computation [204, 93]. Specifically, a neuromorphic chip offers a silicon structural substrate that requires co-design of the software instantiating neuromimetic information processing and the physical hardware that executes such routines. These chips arrange the computation such that processing, and memory are collocated — i.e., in memory processing — thereby eluding the von Neumann bottleneck and offering fault tolerant, sparse parallel and noise-robust computation [304]. Note that multiple chips may be interconnected, allowing the construction of complex circuits that embody functionality such as associative memories or motor control.

Notably, the design of a neuromorphic chip, e.g., [95], is influenced by biophysical models of neuronal activity [153], i.e., spiking neurons. Since biophysical neuronal models can emulate the properties of the neuromorphic hardware that they will be executed on, the use of these chips represents a natural entanglement of physical instantiation with software specification, in the context of co-design *in silico*. Specifically, the inference, learning, and selection are directed by the neuromorphic substrate itself. If one further examines the information processing that underpins networks of biophysical neurons, i.e., spiking neural networks [225], one notices that the inference of a common type of neuron, the leaky integrate-and-fire (LIF) unit, operates according to sets of ordinary differential equations that model the electrical properties of neuronal cells. This neuronal inference influences the plasticity incoming/outgoing synaptic connections i.e., inference drives spike-timing-dependent plasticity. Obviously, the resistance values of the neuromorphic chip's synapses, implemented as memristors [183, 330], affect the effectiveness of each LIF's inference; thus, capturing the inference and learning aspects of MILS.

One way to conceptualize the morphogenetic aspect of MILS in neuromorphic computing is to consider the synaptic connectivity patterns that emerge over time. Specifically, if the model employs a form of local plasticity, e.g., a form of synapse-dependent long-term potentiation (LTP) and depression (LTD) (see supplementary material for a review of key plasticity alternatives), then, as two sets of connected neurons evolve a sparse connectivity structure emerges [51]. Specifically, pre-synaptic neurons that rarely fire in correlation with the spike emission behavior of post-synaptic specializations become disconnected. Since synaptic values are naturally bound to a non-negative

range — due to the properties of the hardware — once a synapse has decayed sufficiently, it is effectively "pruned away". Thus, from the perspective of (synaptic) selection and MILS, a neuromorphic system comprising biophysical LIFs, and local synaptic plasticity, performs a simple kind of morphogenesis, specifically the elimination of redundant synapses. One open question is how the other side of neural morphogenesis, i.e., synaptogenesis and neurogenesis, is instantiated. However, self-replication is still missing in this account: an element that might be fruitful to consider when designing the next generation of neuromorphic mortal computers.

Despite their promise as mortal computers, neuromorphic systems still do not qualify as base-level MCs, because they do not specify base-level imperatives or homeostatic regulation that would drive the processing to assure persistence. Nevertheless, given how close neuromorphic systems are to our definition overall, e.g., instantiating parts of MILS, specification of both software and hardware/substrate, and a design that is crucially morphic, we may be able to modify the information processing of LIF systems to be homeostatic. The interaction of the morphic structure offered with neuromimetic schemes, such as predictive coding and forward-only learning [31, 254, 247] (see supplementary material for details), could provide viable pathways to constructing *in silico* neural-based mortal computers that navigate their environments and forage for resources to charge their neuromorphic batteries.

**The Body-Niche Problem for Mortal Computation.** As we have established throughout this work, a mortal computer needs to distinguish itself from its environment while coupling to it in order to persist. Interestingly, much as animals have adapted cognitive functions that are suitable for their particular niches [34] — e.g., the insect boatman that breathes underwater by trapping air bubbles using the tiny hairs in its abdomen [103] — mortal computers will likely exhibit similar diversity. The algorithmic processes implemented by their physical substrates would constitute the "life cycle", the setting in which it is deployed would constitute its "niche", and its identity would consist of its internal states (their dynamics and relationships, the processes that govern the regulation of life-sustaining dynamics and variable measurements), and its structural boundary would be provided through its Markov blanket. In terms of the niche, the framework of mortal computing hints at a re-framing of the machine intelligence evaluation paradigm: we should be crafting benchmark environments that model resources and evaluate if the agent is able to persist. This is where the free energy formalism, and active inference, offers the greatest promise. If we couple this with designing bodily/morphological systems grounded in real-world principles and dynamics, we might be able to make substantive advances in the design of mortal computers. Additionally, the careful design of the right base-level inductive biases (priors) will be necessary given that successful organisms are the product of a selection history, when viewed through autopoietic enactivist and life-mind continuity lenses.

Among the many problems that researchers in biomimetics and bionics engage with will be how MCs incorporate elements of their niches into an ongoing quest to preserve their internal functional and structural integrity, i.e., how will such systems incorporate the self-orchestrated elements of their Markov blankets and bodily components as well as environmental elements to resist the constant threat of thermodynamic disintegration. The emphasis on a system's environment requires a thorough consideration of both the substrate and niche in which MCs will be instantiated. These aspects will need to be foregrounded; this is a natural consequence of mortal computing's appeal to autopoietic enactivism, embodiment and basal cognition, and the good regulator theorem and law of requisite variety. Potentially, there may be practical difficulties for biomimetics and bionics research efforts to surmount; depending on the choice of substrate and niche — as mentioned, despite their promise, organoids, neuromorphic/photonic platforms, and robots are not readily nor widely accessible or affordable. The same issue, though to a lesser degree, would plague access to various real-world environments. The physical, real-world resource availability constraints inherent to this problem will need to be addressed and solutions will not likely come easy.

To address access to real-world hardware/substrate, which can be prohibitively costly, a possible way to catalyze research and development in mortal computing — and to democratize it — might be the construction of simulated or *virtual morphologies*. For example, in service of crafting complex MCs it might be useful to consider designing subsystems such as digital endocrine (and its biochemical effects) or nervous systems, which are integrated with soft robotics. This means the design of body emulators will need to consider complexity and fidelity depending on the target environment to be navigated, or indeed constructed. Consequently, the separation between the computational intelligence and the system in which they are embedded becomes blurred. One might argue that we should then first build better exoskeletons/bio-cybernetical body systems before we deal with the intelligence *per se.* Nevertheless, the foundations of mortal computing provide a readily available counterpoint: intelligent processing is inexorably intertwined with the body/morphology; again, this is the central thrust of the embodiment thesis and the premise of basal cognition. It will likely be the joint, mutually informed development of the computational cognitive processes, the sensorimotor systems, and the self-regulatory systems that will result in the long-sought after AGI.

The same solution will also apply to environments, although, fortunately, there is a much longer tradition in machine intelligence research of simulating and abstracting such environments, i.e., as in reinforcement learning [239, 203], thus providing some viable starting points for setting up body-niche setups. In the domain of (neuro)robotics [79], viable tools have emerged over the past years [115, 366, 226], offering improved physics engines for simulating real-world robotic systems and some of the task environments in which they would operate. With respect to neuromorphic substrates, simulators of neuromorphic chips [196, 217] and emulator kits, e.g., the RISP neuro-processor [282, 281, 130], have recently emerged; offering an alternative to expensive, non-mainstream hardware. However, while digital emulation might prove to be an invaluable solution, even serving as a force behind the rapid prototyping of early mortal computers, there is always a prescient gap between the fidelity and realism of a particular simulation and the real-world niche that it attempts to emulate, i.e., the "sim2real" problem [173].

One possible guide for designing useful simulated environments might lie in suggested forms of analysis and experimental design in cognitive psychology. For example, in [353], four concrete steps are prescribed for studying embodied cognitive agents: **1)** specify a task (from a first-person perspective) for a perceiving-acting agent, **2)** explicate task-relevant resources available to the agent (spanning the agent's world, body, and brain), **3)** identify how the agent could assemble/use these resources to accomplish the task, and **4)** formulate a means of testing if the agent is using the ensuing solutions based on its performance. Such a multi-step procedure readily provides a clear, goal-directed process for crafting niches for evaluating the functionality of a simulated MC – a setup that promotes problem solving based on the entity's world as well as its morphology, i.e., what are the agent's affordances. This design process could be amended to further incorporate life-critical resources, such as simulated temperature in particular regions of the environment or availability of material sustenance to prioritize metabolic regulation through homeostatic/allostatic processing.

### A.4  Biological Credit Assignment's Interaction with Mortal Systems

From the perspective of our framing of mortal computing, we foresee the gradual rise in efforts in brain-inspired computing and biomimetic credit assignment over the last few decades [165, 202, 242, 171, 300, 215, 263, 264, 259, 261, 298] to become even more important, especially in their generalization and application to spiking neural networks [225]. One of the central goals of biomimetic credit assignment research is to develop mechanisms, or "rules", for adapting the synaptic parameters of neural systems; such mechanisms, ideally, could serve as viable replacements for the core workhorse algorithm of DNNs today – backpropagation of errors (backprop) [221, 294]. In tackling this grand challenge, approaches generally draw inspiration from ideas/results in computational neuroscience and cognitive science to build mechanisms that try to, among other criteria, adjust synaptic weight values using only local information, in both space and time. Furthermore, these approaches require that the same set of mathematical transformations or neural mechanisms be used to carry out both inference and learning; unlike backprop, where the learning step uses mechanisms entirely different those used in inference, i.e., derivative operators to utilize reverse-mode differentiation) in a parallelized fashion. Generally, these approaches embody one or more of the desired criteria but come with tradeoffs, such as being expensive to simulate.

However, several promising schemes have emerged from this line of inquiry including spike-timing-dependent plasticity [225, 51], a form of Hebbian adaptation [165] that applies to biophysical models of neurons, modern forms of contrastive Hebbian learning [301] as well as their spiking neuron variations [231], and forward-only and forward-forward learning [170, 262, 201] and its generalization to the spike-level [255]. Directly relevant to the free energy principle is that of predictive coding [288, 133, 259], including schemes that adhere to/derive similar principles [215, 251, 265], which is a form of neuronal adaptation that learns a generative model by minimizing a variational free energy functional [134] and is a mechanistic scheme that has completely embraced the entanglement of inference and learning historically; [257] originally referred to predictive coding-like adaptation as a form of entangled learning-and-inference.

Within the context of mortal computing, of the growing plethora of brain-inspired algorithms that have emerged over the past several decades, we highlight predictive coding (PC), forward-forward learning (FFL), spike-timing-dependent plasticity (STDP) as some of the initial viable candidate biomimetic credit assignment schemes for satisfying the inference and learning aspects of the MILS property of mortal computing's main definition.

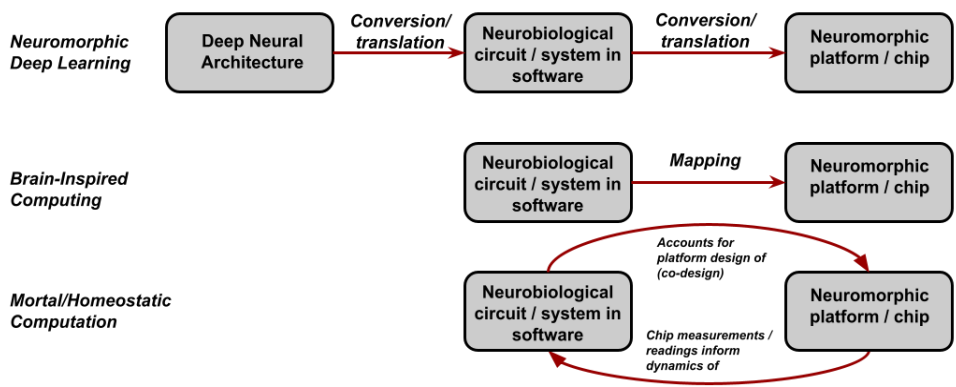

**Figure 8:** The spectrum of brain-motivated computing, ranging from conversion of deep artificial neural networks to a chip/platform to brain-inspired computing and its mapping to a chip to mortal (homeostatic) computing and its direct, entangled relationship with a particular chip/platform.

In terms of what is useful and amenable to integration into mortal computing design is:

- First, all three leverage a system's computational architecture[27] to carry out localized synaptic plasticity in concert with iteratively adjusted neural activity dynamics, at relatively low simulation costs. For instance, PC minimizes local prediction errors in a heterarchical neural message passing structure, FFL systems optimize a set of functionals that contrast positive from negative activity produced by the neural structure, and STDP produces correlational Hebbian updates based on pre- and post-synaptic activity for any synapse within a neural circuit.
- Second, each can be shown to follow the gradient flow induced by a free energy functional; see [133, 134, 298] for details of PC does so, see [262] for details on how a recurrent predictive form FFL does this, and see [43] for STDP's approximate energy function.

Furthermore, PC, FFL, and STDP all (can) operate with fine-grained neuronal dynamics at the spike-train level, facilitating their implementation in neuromoprhic and analog systems. STDP fundamentally inherently does this given that it directly applies to spiking action potentials while PC and FFL can be made to do so; see [255] for details related to PC, see [254] for details related to FFL.

While most current research studies related to these credit assignment schemes develop just the inference (I) and learning (L) elements of mortal computing's MILS, leaving selection of structure (S) largely untouched or unaccounted for, we remark that some efforts have mixed schemes, such as STDP, with evolutionary processes simulated with genetic algorithms. As a result, it would not be inconceivable to specify a morphogenetic alteration process that applies to the computational architecture and software arrangement of neurons, although how this would translate to structurally-adaptive and self-replication mechanisms in actual physical substrate, e.g., neuromorphic chips, that might use algorithms such STDP, FFL, or PC is not clear. Beyond the engineering difficulty inherent to constructing "evolving physical substrates", a good deal of work will need to be done to properly study how schemes such as PC, FFL, and STDP interact with different silicon and possibly biophysical material structures, such as xenobot or organoid constructs. Furthermore, despite standing on a growing body of work that has demonstrated their promising generalization abilities and uniquely useful properties, all three schemes have their issues/drawbacks and largely been tested and studied on amorphic software models that, if implemented, would detrimentally complicate the design of effective mortal computers. Some of these notable issues include:

- STDP is a purely correlational learning scheme, requiring another complementary modulation process, e.g., dopamine/reward signal triggered plasticity [184], to be coupled to it in order for it to carry out predictive/action- oriented tasks, and it requires extremely careful tuning of its biophysical hyperparameters to work well;
- PC can incur expensive iterative inference costs, requires feedback synapses of some form to facilitate any meaningful form of message passing, and a poor design of its feedback structure can complicate

---

[27] Note that this is not the same as a physical morphology.

learning or result in instability [257, 264]; and

- FFL requires a meaningful and useful negative activity/data generating process, even for its non-recurrent, non- structure committing form [170] – how such a process takes form in the context of complicated spiking dynamics poses a significant challenge to overcome [262, 254].

Future work that might incorporate these schemes into physical systems will require addressing these problematic concerns to develop solutions that might leverage the target substrate/morphology. Given that the above schemes have largely been designed and evaluated in the context of amorphous models[28], it will be of greatest importance to craft their generalizations to specific morphologies (an organism vehicle or body) in the context of an environment.

Despite the slower (as compared to backprop-centric deep learning research [298]) yet positive progress being made in biomimetic credit assignment, and despite their viability for use in actual morphological computing constructs as we argued above, these schemes would only serve as one possible tool or aspect in the design of a mortal computing models. The Markov blanket formalism we developed in Section 4.2 might prove helpful in setting the general, broad strokes as to the structure they need to be integrated with and a particular system-niche choice, e.g., a neurorobotic quadruped with the "brain" controller implemented with hierarchically arranged set of memristor crossbars (this is its body morphology) that must find energy charge stations in a warehouse (this is its niche), will help to ground the experimental setup in a proper mortal computing context. In Figure 8, we present several different approaches to examine how learning and inference interact with neuromorphic substrate/hardware. Finally, it might be that the most promising pathway is to not map or adapt current biomimetic credit assignment schemes to a physical substrate but to, instead, formulate completely novel, substrate-specific learning/inference schemes that are grounded in whatever morphogenetic capabilities that may be supported by or could be integrated into the hardware. This is depicted in the bottom row of Figure 8, i.e., "Mortal/Homeostatic Computation". What such a class of biomimetic credit assignment processes would look like is not straightforward. What is clear, however, is that they would be diverse and fundamentally dependent on the body-niche context that they were developed for. In other words, the learning and inference mechanisms for a cellular-like mortal computer moving along chemical concentrations, a bacterium-like agent or xenobot, would be very different from an animal-like mortal computer, such as one in a complex neurorobot operating in mountainous terrain.

## A.5 A Sketch of a Minimal Model of Mortal Computing

Drawing inspiration from the minimal model of metabolism-based control from [111], a next step towards instantiating viable forms of mortal computation would be to possibly craft minimal models of it. In particular, one could begin by developing simulation models of base-level MCs, abstracting several desirable sensorimotor functions. In this section, we construct a simple theoretical sketch of what such a minimal model/agent might look like, demonstrating a rudimentary homeostatic regulatory process (based on energy level) and how such an agent's internal, Markov blanket (including its action/sensing/energy exchange), and external states could be specified for simulation.

**Mortal Predictive Coding: Energy-Aware Computation.** Inspired by the physiological notion of hunger and its relation to energy balance [213], we adapt the concept to a simple neuro-mimetic circuit, an instance of what we label as "mortal predictive coding". The homeostatic control process of its energy balance pertains to the amount of energy taken in through nutrition needed for an organism to match the amount of energy it uses (to conduct work). This gives rise to the concept of "appetite" which is biophysically regulated by two key hormones – grephlin, which stimulates hunger and the intake of food/nutrients, and leptin, which signals satiety or fullness [250]. We could vastly simplify these hormonal signifiers and relate them to a measurable energy quantity commonly found in most robotic systems: "battery level" or "power level".

Our computational organism (see Figure 9), which we label in this sketch as a "homeostatic predictive coding circuit" (H-PCC), is characterized by the following three non-neural bodily components: **1)** a bodily membrane that governs the rate and quantity at which energy is absorbed by the system's internal battery (this could be simply specified as an ordinary differential equation characterizing energy exchange between the H-PCC organism and its world), **2)** a battery sub-module that is wired to a set of system quality control states (a simple setup would be one state per discretized power level of the battery), **3)** a set of sensory nodes to perceive a nearby portion of the system's world (such as an ego-centric view of the pixels in a small region around the front of the body), and **4)** a set of actuators that facilitates the system's movement (in a simplified environment, these could be fixed movements in particular cardinal directions) as well as one that couples, or "plugs in", the system to a charging port, each associated with a particular cost that will drain the battery. Note that these elements provide a basic description of the H-PCC's Markov blanket and further characterize a core internal state, the battery state, that will serve as the basis for a homeostatic regulation. In other words, the system will need to keep the battery sufficiently charged in order to compensate for the energy costs of its actions. A base/non-neural homeostatic mechanism would further be implemented to possibly drain any of the system's energy stores so as to prioritize keeping the battery's level above a minimum threshold value).

---

[28] With the exception of STDP, which was only later applied to amorphous constructs after emerging from the study of actual biological neurons and how synapses adjusted themselves biochemically [51, 52].

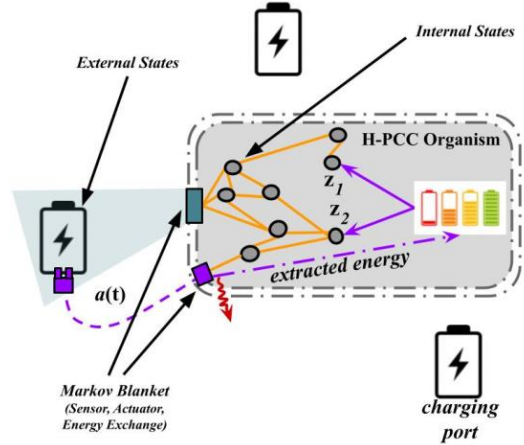

**Figure 9:** A depiction of the theoretical example of the homeostatic predictive coding circuit (H-PCC).

With respect to its internal states, each node or cell in the neural portion of the system would be assigned a particular physical type. For instance, each node is a multi-compartment neuron with at least one portion dedicated to maintaining the neuron's current temporal state such as its voltage membrane potential, expanded to account for a particular physical area/volume, and another dedicated to receiving a modulatory input from a projection of the battery module's control states. Some cells, closer to the H-PPC's bodily actuators, would be wired directly to output commands. Each neural cell is further conditioned by the agent's previously chosen action $a(t-1)$. Learning would be conducted within the framework of spiking predictive coding [255] (note that morphogenesis could take the form of a synaptic generation, such as synaptogenesis, in tandem with a synaptic pruning algorithm). The H-PCC's environment, or external states, would contain several battery charging ports that this system may couple to, but only for a finite amount of recharge over a finite time duration, and the agent's goal will be to keep itself charged over a far longer time-horizon. A specification will be needed for the thermodynamics of how energy is utilized by the agent's bodily actuators to perform work as well as how energy leaves the body to the niche and becomes a heat-like quantity.

The H-PCC must optimize its variational free energy $\mathcal{F}(Z, b)$ where $Z$ is the set of battery control states $\{z_1, z_2, \ldots, z_B\}$ and $b$ is the negative of the battery reading (lower is better). One would then likely decompose $\mathcal{F}(Z, b)$ into a perceptual free energy term, e.g., full hierarchical Gaussian log likelihood, and a heavily weighted battery term. If the battery term $b$ reaches/tends to 0, the H-PCC "dies". The H-PCC continues to maintain its homeostasis with respect to $b$ so long as it minimizes its free energy, finding states of the world where $b$ is high. Finally, in our theoretical sketch, one could envision designing planning schemes based on rolling out possible futures given that neural regions are conditioned on previous decisions – this would induce more sophisticated action sequences that possibly lead to behavior that finds portions of the environment with reliable charging ports to plug into, seeking favorable higher battery states.

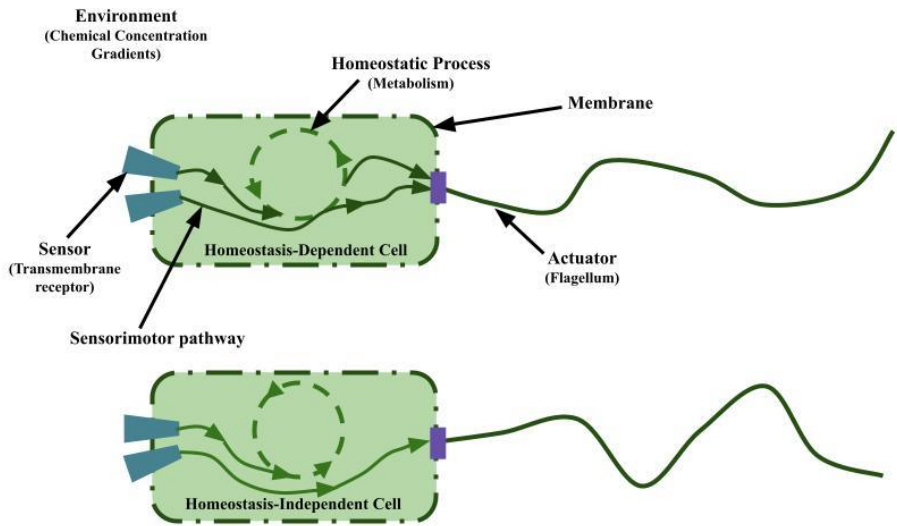

**Figure 10:** A visual depiction of two different ways that computation is carried out in E. Coli cell (our figure is inspired by the ones in [111]) – the top model presents a cell where its sensorimotor pathways are modulated by its internal homeostatic process(es) and its bodily actuator (its flagellum) is further directly driven by its metabolism, a metabolism (homeostasis) dependent/based cell. The bottom cell model depicts the traditional/classical model of E. Coli as a metabolism (homeostasis) independent system, i.e., its movement/actuator are purely reactive and driven by sensory information only. Note this cell model could be considered to be a possible instantiation of a mortal computer, i.e., its body consists of the cell membrane, the flagellum, and the transmembrane receptors), hence our use of the label "homeostasis" which is motivated by the fact that homeostasis regulates metabolic processes/pathways and could be argued to a core driver of a "living agent's" actions.

## A.6    Homeostatic/Metabolic Dependency: A Chemotaxis Example

In the biophysics section, the main paper established as part of its characterization of a mortal computer that such a system would implement sensorimotor control sub-systems that are dependent in various ways on its regulatory sub-systems. This type of dependency is inspired and motivated by work on metabolic processing in biology, particularly the strain of work that has advanced the concept of metabolic (metabolism) dependency. Concretely, metabolism dependency is the driving force behind a cell's motility, or its independent motion carried out due to metabolic energy, such as that of the well-studied unicellular entity known as Escherichia coli (E. coli). A bacterium, e.g., E. coli, implements a strategy for moving up (up-gradient) or down (down-gradient) an attracting substance concentration gradient by rotating its flagella in one of two directions. If the cell rotates its flagella counterclockwise, it will exhibit a directed motion (labeled as "running"), whereas if it rotates its flagella clockwise, it will exhibit (rotational) motion (referred to as "tumbling") [349]. E. coli, which internally compares current and past concentrations of attractants, runs if an attractant's concentration is increasing and tumbles otherwise, i.e., this is also known as the process of "adaptive gradient climbing". Alternatively, the bacterium combines tumbling with running to produce a random walk until certain (chemical) concentrations are high enough such that, eventually, tumbling becomes the dominant motion (allowing it to remain in place), gradually phasing out the running action, i.e., the process of "selective-stopping" [112].

Historically, studies of the chemotaxis, or movement of cells/organisms in the direction of the gradient of increasing/decreasing concentrations of a substance, of bacteria, such as the running and tumbling of the E. coli cell described above, has held the prevailing view that the cell's short-term activity is independent of metabolism, i.e., it is metabolism-independent [10]. More specifically, the chemotaxis of a bacterial organism is not influenced by its current metabolic state and only influenced by its environment (specifically, it is only responsive to the concentration of attraction chemicals [237, 349]) – a cell's movement is not driven by the effect of the (chemical) attractant upon its metabolism but exclusively by the way that attractants stimulate/excite its transmembrane sensors. In essence, although the organism's sensors, actuators, and transduction processes are the product of its metabolism, they are not modulated or effected by it. In contrast, an alternative to this view is that of metabolism-dependent chemotaxis [88, 324, 12], which has begun to re-emerge and posits that metabolism itself has an ongoing influence on the cell's behavior and is key driver behind its adaptability [111]. In other words, metabolic processing centrally affects the organism's contextualized, integrative evaluations of its environment – simply stated, the cells are capable of modifying themselves in response to environmental changes. See Figure 10 for a generalized visual depiction chemotaxis from the point-of-view of metabolic independence versus dependence. In addition, evidence in support of (dominant) metabolism-dependent behavior has been steadily growing for a variety of bacteria species [188, 13, 158, 299], including E. coli itself [367, 325]. Beyond challenging long-standing results and long-held assumptions in the biological study of bacteria and unicellular organisms [33, 267, 156], metabolic dependent processing has been increasingly argued to be a definite modulator of, and even key driver the evolutionary process behind, behavioral and strategic processes that could underpin basal cognition [224].

Notably, [111] introduced a version of metabolic dependency in chemotaxis that they labeled as metabolism-based chemotaxis. Unlike its metabolism-sensitive predecessor (where flagellar rotation is coupled with / influenced by metabolic processing), metabolism-based chemotaxis claims that metabolism plays a central role in directly modulating the cell's behavior itself. [111] presented a minimal computational model of metabolism-based chemotaxis and demonstrated, via simulation, a wide range of effects that aligned with the expected behavioral patterns of real E. coli cells. From this paper's perspective, the minimal model of [111] could serve as an inspiring prototype template for mortal computation itself.

While the minimal metabolism model did not account for specific biological mechanisms, it offered a key proof-of-concept that a simple abstraction/model of metabolism could support a wide and varied range of observed aspects of chemotactic functionality. This has sparked a reconsideration of bacterial organisms as exclusively stimulus-driven systems [12, 244, 152] – such living systems should be treated as ones where their attractants/repellents should be framed in terms of metabolism, influenced by the cell's history and internal organization (metabolic rates, active/non-active metabolic pathways, etc.). For example, the process theory of active inference [140] has been applied to implement action-oriented dynamical models [32] of phototaxis [187]. In general, metabolism dependent and metabolism-based behavior in living systems places metabolic drive as the central impetus behind an organism's interaction with its niche. This metabolic drive and environmental factors co-exist in the picture of organism adaptation and behavior meaning that life, in general, has a metabolic side to it [59]. Note that, in the main paper, we generalized the concept of metabolic dependency to notions including homeostatic and allostatic dependency in order to emphasize different levels of modulation on an organizationally closed system's perception and decision-making.